\documentclass[prl,twocolumn,letterpaper, superscriptaddress]{revtex4-1}
\usepackage{graphicx}
\usepackage{dcolumn}
\usepackage{bm}
\usepackage{color}
\usepackage{tabularx}
\usepackage{array}
\usepackage{amsmath}
\usepackage{stmaryrd}  

\begin{document}

\title{Probing phase coupling between two spin-torque nano-oscillators\\ with an external source}

\author{Yi Li}
\email{yi.li@cea.fr}
\affiliation{Service de Physique de l'\'{E}tat Condens\'{e}, CEA, CNRS, Universit\'{e} Paris-Saclay, Gif-sur-Yvette, France}

\author{Xavier de Milly}
\affiliation{Service de Physique de l'\'{E}tat Condens\'{e}, CEA, CNRS, Universit\'{e} Paris-Saclay, Gif-sur-Yvette, France}

\author{Flavio Abreu Araujo}
\affiliation{Unit\'{e} Mixte de Physique CNRS, Thales, Univ. Paris-Sud, Universit\'{e} Paris-Saclay, Palaiseau, France}

\author{Olivier Klein}
\affiliation{SPINTEC, Univ. Grenoble Alpes / CEA / CNRS, 38000 Grenoble, France}

\author{Vincent Cros}
\affiliation{Unit\'{e} Mixte de Physique CNRS, Thales, Univ. Paris-Sud, Universit\'{e} Paris-Saclay, Palaiseau, France}

\author{Julie Grollier}
\affiliation{Unit\'{e} Mixte de Physique CNRS, Thales, Univ. Paris-Sud, Universit\'{e} Paris-Saclay, Palaiseau, France}

\author{Gr\'{e}goire de Loubens}
\email{gregoire.deloubens@cea.fr}
\affiliation{Service de Physique de l'\'{E}tat Condens\'{e}, CEA, CNRS, Universit\'{e} Paris-Saclay, Gif-sur-Yvette, France}

\date{\today}

\begin{abstract}

Phase coupling between auto-oscillators is central for achieving coherent responses such as synchronization. Here we present an experimental approach to probe it in the case of two dipolarly coupled spin-torque vortex nano-oscillators using an external microwave field. By phase-locking one oscillator to the external source, we observe frequency pulling on the second oscillator. From coupled phase equations we show analytically that this frequency pulling results from concerted actions of oscillator-oscillator and source-oscillator couplings. The analysis allows us to determine the strength and phase shift of coupling between two oscillators, yielding important information for the implementation of large interacting oscillator networks.

\end{abstract}

\maketitle

Self-sustained oscillators which are linked by phase coupling exhibit abundant collective dynamics \cite{PikovskyBook2001} and describe diverse systems in nature \cite{WiesenfeldPRL1996,ShimScience2007,HeinrichPRL2011,MZhangPRL2012,KakaNature2005,MancoffNature2005,KissScience2002,TinsleyNphys2012,HartwellNature1999}. In particular, they can synchronize, which is important in the fields of engineering, biology and computing. Indeed, synchronized oscillators exhibit improved amplitudes and spectral purity of their outputs, and can be used to study and mimic neural networks \cite{LocatelliNmat2013,GrollierProcIEEE2016}. Theoretical explorations of this phenomenon have been ongoing for decades in particular within the framework of the Kuramoto model \cite{KuramotoBook1984,AcebronRMP2005}, where phase coupling is simplified as a sinusoidal function of phase difference:
\begin{equation}
{d\varphi_i \over dt} = \omega_i + \sum_{j}\Omega_{ji}\sin(\varphi_j - \varphi_i + \beta_i)
\label{eq0}
\end{equation}
where $\varphi_{i}$ is the phase of $i$th oscillator, $\omega_i$ is its free-running frequency, $\Omega_{ji}$ is the coupling strength between $j$th and $i$th oscillators and $\beta_i$ is an intrinsic phase shift related to the nature of the coupling and to the nonlinearity of the oscillator \cite{SlavinIEEE2009}. In experiments, technological progress has allowed mutual synchronization in many systems compatible with lithographic fabrications, such as Josephson junctions \cite{WiesenfeldPRL1996}, nanomechanical and optomechanical structures \cite{ShimScience2007,HeinrichPRL2011,MZhangPRL2012}, and spin-torque nano-oscillators \cite{KakaNature2005,MancoffNature2005,RuotoloNnano2009,SaniNcomm2013,locatelliSREP2015,LebrunArXiv2016}. The strength of synchronization in all these systems is set by the coupling parameters in Eq. (\ref{eq0}). However, the coupling  strength $\Omega_{ji}$ and the intrinsic phase shift  $\beta_i$ are rarely quantified in experiments despite their importance for achieving large phase-locking ranges \cite{TiberkevichAPL2009,OlehPRL2012}. Being able to quantify these parameters is also crucial for synchronization-based information processing such as coupled-oscillator associative memories \cite{csaba12,nikonov15}.

Among different oscillator systems, spin-torque nano-oscillators \cite{KiselevNature2003} serve as outstanding candidates for implementing coupled oscillator arrays, due to their sub-micron dimensions, nonlinear behaviors with large frequency tunability, simple signal extractions from magnetoresistance and ease to be coupled and synchronized \cite{KakaNature2005,MancoffNature2005,RuotoloNnano2009,SaniNcomm2013,locatelliSREP2015,LebrunArXiv2016,RippardPRL2005,GeorgesPRL2008,UrazhdinPRL2010,QuinsatAPL2011,HamadehAPL2014,LebrunPRL2015}. Of special engineering interests are spin-torque vortex oscillators \cite{PribiagNphys2007,DussauxNcomm2010} which allow operation without biasing field and different tuning properties \cite{LocatelliAPL2011,HamadehPRL2014,SlukaNcomm2015} linked to the bistable orientation of the vortex core magnetization (polarity) \cite{deLoubensPRL2009}. The synchronization of two adjacent vortex oscillators through their dipolar field \cite{ShibataPRB2003,SugimotoPRL2011,HanSREP2013} has been demonstrated \cite{BelanovskyPRB2012,BelanovskyAPL2013,locatelliSREP2015,AbreuAraujoJAP2016}, as well as the control of the phase-locking bandwidth by their relative vortex polarities \cite{AbreuAraujoPRB2015,locatelliSREP2015}. Moreover, vortex oscillators are a model system for coupled oscillators in general, because their dynamics is well understood \cite{DussauxPRB2012} and their phase-coupling can be described by Kuramoto-like equations \cite{locatelliSREP2015,FlovikSREP2016}.

In this work we employ a third reference ``oscillator'', namely, an external microwave field with tunable frequency and power, as a dynamical probe to measure the dipolar coupling between two spin-torque vortex oscillators. When the microwave field phase-locks one oscillator, an obvious frequency pulling is measured on the second oscillator. By including the coupling to external source in Eq. (\ref{eq0}), we show analytically that this frequency pulling is due to in-phase actions of source-oscillator and inter-oscillator couplings within the phase-locking bandwidths, beyond which it disappears. The model is tested upon varying the source-oscillator coupling by changing the microwave power, as well as the inter-oscillator coupling by changing the vortex polarity states. It allows to extract dipolar coupling strengths and phase shifts, with the former compatible with analytical calculations \cite{AbreuAraujoPRB2015}. Our results provide a new way to directly reveal and characterize the mutual coupling between oscillators through their attraction to a third reference oscillator, which can be applied to various oscillator systems.

\begin{figure}[htb]
\centering
\includegraphics[width=3.0 in]{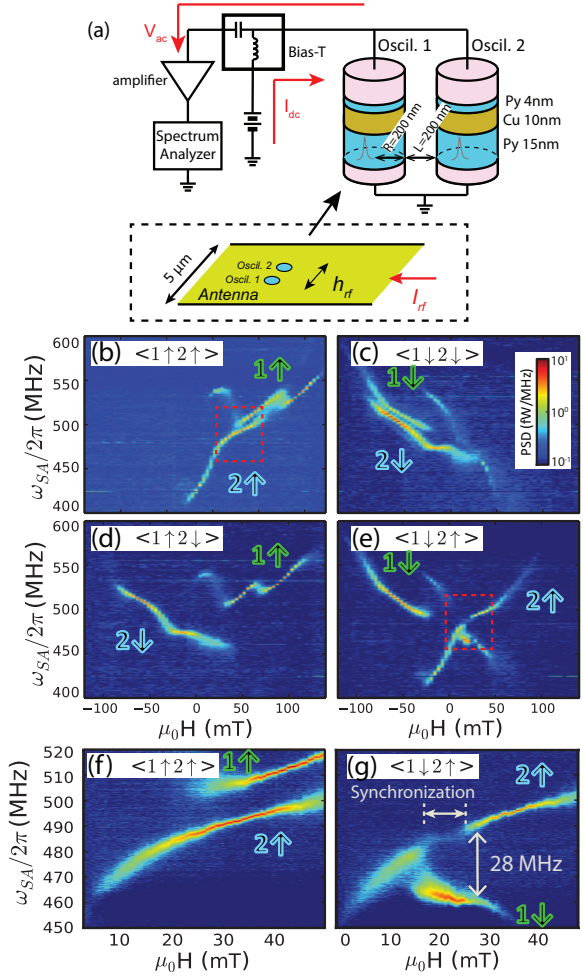}
\caption{(a) Schematics of the sample and electrical circuit. (b-g) Power spectral density maps of auto-oscillation modes in log scale. Four different polarity states of the two thick Py vortex layers are shown: (b)$<1\uparrow 2\uparrow>$, (c)$<1\downarrow 2\downarrow>$, (d)$<1\uparrow 2\downarrow>$  and (e)$<1\downarrow 2\uparrow>$. (f,g) Zoomed-in power spectral density data of $<1\uparrow 2\uparrow>$ and $<1\downarrow 2\uparrow>$ for the red box regions of (b) and (e), respectively. Mutual synchronization is observed between $\mu_0 H=17.5$ and 26.0 mT in $<1\downarrow 2\uparrow>$ state.}
\label{fig1}
\end{figure}

Our sample consists of two cylindrical spin-torque nano-oscillators with identical nominal diameters of $2R=400$ nm and an edge-to-edge separation of $L=200$ nm, as shown in Fig. \ref{fig1}(a). Each oscillator has a spin-valve layer structure of Py(15 nm)/Cu(10 nm)/Py(4 nm) (Py = Ni$_{80}$Fe$_{20}$). During operation a strong dc current is injected through the two oscillators in parallel, which favors a vortex state in all Py layers \cite{supplement}. It flows from Py(4 nm) to Py(15 nm), so that in each oscillator the spin-transfer torque destabilizes the mode dominated by the thick layer and overdamps the thin layer dynamics \cite{KhvalkovskiyAPL2010,LocatelliAPL2011,HamadehPRL2014}. The current is set to 95 mA, i.e., 1.5 times as the critical current to drive the auto-oscillations of the thick Py layers. The dynamics excited in each oscillator corresponds to the rotation of the vortex core around its equilibrium position, the so-called gyrotropic mode \cite{GuslienkoJAP2002}. Because of the small lateral separation, the two oscillators are dynamically coupled through their dipolar field \cite{BelanovskyPRB2012,locatelliSREP2015}. We note that owing to their much smaller volumes and limited dynamics, the contribution of the thin layer vortices to the oscillator-oscillator coupling is weak. Moreover, their vortex core polarity is not purposely controlled in this study. In the following, we will thus refer exclusively to the vortices in the thick Py layers of each oscillator, labeled 1 and 2. To provide an external rf field, an electrically isolated antenna is patterned on top of the sample \cite{NaletovPRB2011}, creating an in-plane $h_{rf}$ linearly polarized along the direction made by the two oscillators (see Fig. \ref{fig1}a). Furthermore a biasing magnetic field is applied perpendicular to the sample plane in order to vary the gyrotropic frequencies \cite{deLoubensPRL2009,DussauxPRB2012}.

First we examine the microwave signals associated with auto-oscillations in each oscillator. Figs. \ref{fig1}(b-e) show the color maps of the power spectral density as a function of perpendicular field $H$. In each graph, two branches corresponding to the gyrotropic modes of the thick layer vortex in each oscillator are observed. The four combined vortex polarity states for oscillators 1 and 2 can be obtained after applying well-chosen perpendicular switching fields \cite{LocatelliAPL2011}. The polarity state of oscillator $i$ is defined as $<i\uparrow>$ ($<i\downarrow>$) for vortex core magnetization parallel (antiparallel) to the positive biasing field direction, which corresponds to a positive (negative) frequency-field slope \cite{deLoubensPRL2009,DussauxPRB2012}.

Next we demonstrate the existence of dipolar coupling by the observation of mutual synchronization. Figs. \ref{fig1}(f,g) compare the zoomed-in power spectra of $<1\uparrow 2\uparrow>$ and $<1\downarrow 2\uparrow>$ states for $0\le \mu_0 H \le 50$ mT, as labeled by the red boxes in Figs. \ref{fig1}(b) and (e), respectively. By switching the polarity of vortex oscillator 1, a clear gap of the auto-oscillation branch for oscillator 2 is found between $\mu_0 H=17.5$ and 26 mT in $<1\downarrow 2\uparrow>$ state, while for $<1\uparrow 2\uparrow>$ state the branch is continuous. This gap, accompanied by a bright lower-frequency branch, is associated to the synchronization of the two oscillators. From the right edge of the synchronization bandwidth we deduce that the maximal frequency mismatch for mutual synchronization is 28 MHz. The frequency mismatch corresponding to the unlocking of the two oscillators at the left edge is smaller. We attribute this to the fact that the amplitude and linewidth of oscillators can vary with the perpendicular field \cite{HamadehPRL2014}, which will change the effective dipolar coupling. The results above show that the dipolar interaction is strong enough to synchronize the two oscillators. Still, a quantitative evaluation of its strength $\Omega_{ji}$ and phase shift $\beta_i$ is lacking at this point of the analysis.

\begin{figure}[htb]
 \centering
 \includegraphics[width=2.8 in]{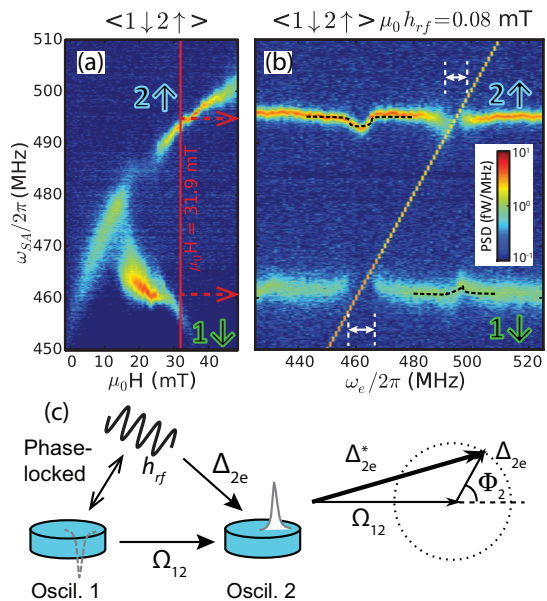}
 \caption{(a) Location of spectra at $<1\downarrow 2\uparrow>$ state for the microwave study. (b) Auto-oscillation spectra as a function of microwave field frequency for $\mu_0 H=31.9$ mT, indicated in (a). Signal from the source appears as the oblique narrow line. The microwave power is $-23$ dBm, corresponding to $\mu_0 h_{rf}=0.08$ mT. White arrows show the phase-locking bandwidths. Black dashed curves are the fits to Eq. (\ref{eq3}). (c) Vector diagram of $\Omega_{12}$ and $\Delta_{2e}$ when oscillator 1 is phase-locked to the microwave field. }
 \label{fig2}
\end{figure}

In order to directly reveal and quantify the dipolar coupling, we fix both the biasing current and magnetic field and apply a weak microwave field, which couples to both oscillators. The two oscillators are set to an unsynchronized state at $\mu_0 H=31.9$ mT, shown in Fig. \ref{fig2}(a). Fig. \ref{fig2}(b) shows the evolution of auto-oscillation peaks of the two oscillators as a function of the external microwave field frequency $\omega_e$. When $\omega_e$ crosses the peak of oscillator 1 around 460 MHz, the disappearance of the peak reflects the phase locking to the external rf source \cite{RippardPRL2005,GeorgesPRL2008,HamadehAPL2014}. In addition, we also detect a significant frequency pulling on oscillator 2. This is a striking observation, because the frequency mismatch between oscillators, $(\omega_2 - \omega_1)/2\pi = 35 $ MHz, is five times larger than the phase-locking bandwidths, around 7 MHz, of the two oscillators to the external source. The remote frequency pulling is a strong indication of coupling between the two oscillators as it is bound to the phase-locking bandwidth. It is important to note that no obvious frequency shift is observed when $\omega_e$ lies between the two auto-oscillation peaks. Reciprocally, a similar effect is also observed on oscillator 1 when oscillator 2 is phase-locked to the microwave field around 495 MHz.

To understand these phenomena, we develop a simplified analytical formalism based on general oscillator equations \cite{SlavinIEEE2009}. For two dipolarly coupled vortex oscillators experiencing a linearly polarized microwave field, the phase equations can be formulated \cite{supplement} from the Thiele equation which describes the vortex core dynamics in a magnetic dot \cite{GuslienkoAPL2006,IvanovPRL2007,KhvalkovskiyPRB2009}, as:
\begin{align}
-&{d\theta_i \over dt} +\omega_e-\omega_{i g} +\Omega_{j i} \cos(\theta_i+\gamma_i^{NL} - \theta_j) \nonumber \\
&- \Delta_{i e}\sin(\theta_i+\gamma_i^{NL}+\gamma_i^{rf}) = 0
\label{eq1}
\end{align}
where $\theta_i=\omega_e t  - p_i\varphi_i$ is the phase difference between the microwave field and the position of the vortex core, $p_i=\pm 1$ is the vortex polarity, $\omega_{i g}$ is the free-running frequency of oscillator $i$, $\Omega_{ji}=\Omega(X_j/X_i)$ is the dipolar coupling strength $\Omega$ normalized by the ratio of vortex gyration amplitudes $X_j/X_i$ and $\Delta_{ie}$ is the coupling strength to the external microwave source. The index is defined as $(i, j)=(1,2)$ or $(2,1)$. In Eq. (\ref{eq1}) two additional phases are present:  $\gamma_i^{NL}$ is the intrinsic phase shift introduced by the nonlinearity of the oscillators \cite{SlavinIEEE2009,TiberkevichAPL2009}; $\gamma_i^{rf}$ is the microwave coupling phase, determined by the geometric alignment of the microwave field to each oscillator (see Fig. \ref{fig1}(a)). We highlight that Eq. (\ref{eq1}) describes the general behaviors of self-sustained oscillators: for $\Delta_{i e}=0$, it is reduced to Kuramoto equations Eq. (\ref{eq0}) with $\beta_i = \pi/2 - \gamma_i^{NL}$, where the $\pi/2$ phase originates from the conservative nature of dipolar coupling \cite{SlavinIEEE2009}; for $\Omega=0$, it is reduced to the Adler equation responsible for one oscillator phase-locking to an external source \cite{AdlerIEEE1973}.

In the general case, the phase dynamics of oscillator 2 evolves in a complex way due to the uncorrelated forces exerted by the microwave field and oscillator 1. However, when oscillator 1 phase-locks to the microwave field, the situation simplifies: its relative phase with respect to the microwave field, $\theta_1$, becomes a constant. In that case, we can rewrite the phase dynamics of oscillator 2 in Eq. (\ref{eq1}) as driven solely by the action of the microwave field, but with a modified effective coupling strength $\Delta_{2e}^*$ that takes into account both the microwave couplings, and the dipolar attraction to oscillator 1:
\begin{equation}
 \Delta_{2e}^* = \sqrt{ \Delta_{2e}^2 + \Omega_{12}^2 - 2\Delta_{2e}\Omega_{12}\cos\Phi_2 }
 \label{eq2}
\end{equation}
From Eq. (\ref{eq2}), we find $\Delta_{2e}^*$ is the vector sum of the effective dipolar coupling strength $\Omega_{12}$ and the microwave coupling strength $\Delta_{2e}$ with a phase difference $\Phi_2=\theta_1+\gamma_2^{rf}+\pi/2$ (Fig. \ref{fig2}c). The frequency of oscillator 2 is then determined by the frequency of the microwave field, and the strength of this new effective coupling $\Delta_{2e}$ through:
\begin{equation}
\omega_2 = \omega_e \pm \sqrt{(\omega_e-\omega_{2 g})^2 - ( \Delta_{2e}^*)^2}
\label{eq3}
\end{equation}
where ``$\pm$'' depends on the sign of $\omega_e-\omega_{2 g}$. Eq. (\ref{eq3}) indicates that when oscillator 1 is phase locked to the microwave field, it can help pulling the frequency of oscillator 2 towards the frequency of the source, as observed in Fig. \ref{fig2}(b).

\begin{figure}[htb]
  \includegraphics[width=3.0 in]{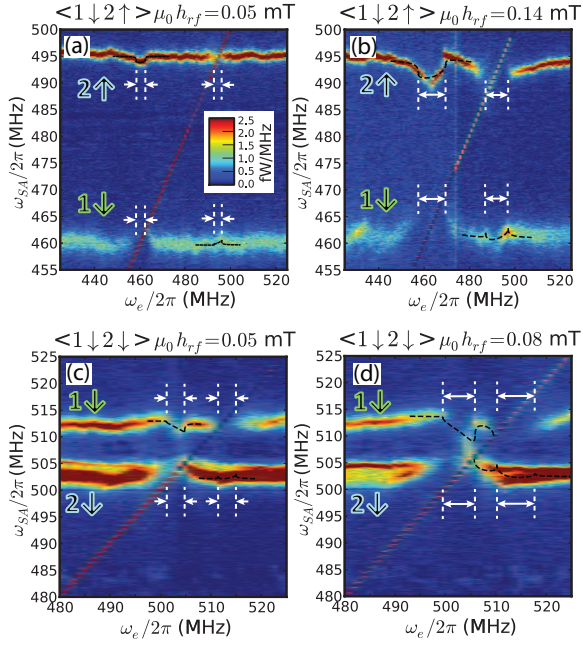}
  \caption{Probing dipolar coupling in various conditions. (a-b) $<1\downarrow 2\uparrow>$ state for $\mu_0 H=31.9$ mT, with microwave powers of (a) $-28$ dBm and (b) $-18$ dBm. (c-d) $<1\downarrow 2\downarrow>$ state for $\mu_0 H=-61.6$ mT with microwave power of (c) $-28$ dBm and (d) $-23$ dBm. White arrows show the phase-locking bandwidths. Black dashed curves are the fits to Eq. (\ref{eq3}). The fit parameters are listed in Table \ref{table1}.}
 \label{fig3}
\end{figure}

\begin{table}[ht]
\centering
\setlength{\extrarowheight}{1.5pt}
\begin{tabular}{>{\centering\arraybackslash}m{0.9in} | >{\centering\arraybackslash}m{0.4in} >{\centering\arraybackslash}m{0.4in} >{\centering\arraybackslash}m{0.4in} | >{\centering\arraybackslash}m{0.4in} >{\centering\arraybackslash}m{0.4in} }
&\multicolumn{3}{c|}{antiparallel} & \multicolumn{2}{c}{parallel} \\ 
\hline
$\mu_0 h_{rf}$ (mT)&0.05 & 0.08 & 0.14 & 0.05 & 0.08 \\
\hline
\hline
${\Delta_{1e}+\Delta_{2e}\over 2\cdot 2\pi}$ (MHz) & 1.8 & 3.2 &5.5 & 1.8 & 3.5\\
$\Omega/2\pi$ (MHz) & -6.7 & -7.9 & -9.3 & 3.6 & 4.2\\
$\gamma_1^{NL}$ (rad) & -2.7 & -2.8 &-2.0 & 2.6 & 2.1\\
$\gamma_2^{NL}$ (rad) & 1.1 & 1.1 & 0.7 & -2.1 & -1.4\\
\hline
\end{tabular}
\caption{Fit parameters of Figs. \ref{fig2}(b) and \ref{fig3}. The values of $h_{rf}$ are calculated from the antenna geometry. The signs of $\Omega$ are fixed to the predictions in Ref. \cite{AbreuAraujoPRB2015}.}
\label{table1}
\end{table}

Full analytical solutions to our model can be obtained in the limit of weak microwave coupling \cite{supplement}. In Fig. \ref{fig3} we use them to extract the coupling parameters under different conditions. First, the microwave power is varied, which sets the phase-locking bandwidths and associated remote frequency pullings. Second, both antiparallel (Figs. \ref{fig3}a,b) and parallel (Figs. \ref{fig3}c,d) vortex polarity alignments are examined, for which the strength of dipolar coupling is expected to change by a factor close to three \cite{AbreuAraujoPRB2015,locatelliSREP2015}. The data are fitted to Eq. (\ref{eq3}) with $\Omega$ and $\gamma_i^{NL}$ as the fit parameters. Positive and negative signs of $\Omega$ are expected for parallel and antiparallel polarity alignments, respectively \cite{AbreuAraujoPRB2015}, which is taken into account. Details about the fitting procedure can be found in the Supplemental Information \cite{supplement}. The fitting curves are shown in Fig. \ref{fig3} and Fig. \ref{fig2}(b).

Table I lists the fitting results along with the microwave field amplitude $h_{rf}$. As expected the mean of phase-locking strengths $(\Delta_{1e}+\Delta_{2e})/2$ is proportional to the microwave field. For the antiparallel polarity alignment, the extracted dipolar coupling $\Omega$ slightly increases with $h_{rf}$. One reason is that the vortex gyration amplitude $X_{i}$ might be increased as oscillator $i$ is phase-locked to the microwave field, resulting in an enhancement of $\Omega_{ij}$ on oscillator $j$. Another possibility is the incomplete phase locking at small $h_{rf}$ (observed in Fig. \ref{fig3}a for oscillator 2) due to thermal fluctuations, which are likely to reduce the effective coupling \cite{GeorgesPRL2008,HamadehAPL2014}. Owing to these two counteracting effects, we take the average of the three experiments, $\Omega_\text{AP}/2\pi= -8.0$ MHz, as the extracted value of $\Omega$. For the parallel polarity state, we take the value extracted from Fig. \ref{fig3}(c), $\Omega_\text{P}/2\pi= 3.6$ MHz, as the strength of the dipolar coupling. In fact, the limit of weak microwave coupling does not hold in Fig. \ref{fig3}(d) due to the large microwave power in comparison to the small frequency mismatch between oscillators, which results in a more complex dynamics. It is interesting to note that the two values compare favorably with the ones obtained from macrodipole approximation taking into account solely the thick Py layers \cite{BelanovskyPRB2012,AbreuAraujoPRB2015}. In that case $\Omega_\text{AP(P)}/2\pi=-3(+1)\cdot\xi^2\gamma \mu_0M_sR^2h/32\pi d^3=-6.9(2.3)$ MHz, where $\xi=2/3$ from the two-vortex ansatz \cite{GuslienkoJAP2002}, $\gamma/2\pi=29.7$ GHz/T is the gyromagnetic ratio, $\mu_0M_s=0.96$ T the saturation magnetization of the Py vortex layers \cite{NaletovPRB2011}, $h$ their thickness, and $d=2R+L$ the center-to-center distance between oscillators. We also point out that the ratio $\Omega_\text{AP}/\Omega_\text{P}$ in our experiment agrees with the ratio of critical frequency mismatch, $\Delta f_\text{AP}/\Delta f_\text{P}= 2.4$ in our prior work \cite{locatelliSREP2015}, and depends on the exact geometry of the oscillator pair \cite{AbreuAraujoPRB2015}.

The phase shift $\gamma_i^{NL}$ is linked to the position of the largest remote frequency pulling in Fig. \ref{fig3}. In the antiparallel polarity alignment the values of $\gamma_i^{NL}$ are reproducible at various microwave fields but differ from that in the parallel alignment, indicating large variations of parameters in magnetic dynamics upon polarity change. From the model, $\tan \gamma_i^{NL}$ is the reduced nonlinear coefficient $\nu_i$ of oscillator $i$ \cite{SlavinIEEE2009}. However we note that the fitting results with negative values of $\gamma_i^{NL}$ point towards either additional extrinsic phase due, \textit{e.g.}, to parasitic rf couplings between the antenna and sample circuits, or more complex dynamics than assumed in the simple analytical model.

One interesting finding is that the extracted $\Omega$ in the antiparallel polarity alignment is much smaller than the phase-locking frequency mismatch of 28 MHz found in Fig. \ref{fig1}(g). In our experiments the amplitude ratio $X_2/X_1$ is close to one \cite{supplement}. In the phase-locking solution derived by Slavin and Tiberkevich \cite{SlavinPRB2006}, the maximal frequency mismatch for mutual synchronization is then $\Omega(\nu_1+\nu_2)$. Thus we confirm the role of nonlinearities, with $\nu_1+\nu_2$ around 3.5, in the large phase-locking frequency mismatch. The fact that the synchronized mode is closer to the peak branch of oscillator 1 likely indicates that $\nu_2$ is greater than $\nu_1$, making it easier for oscillator 2 to adapt its frequency to oscillator 1.

Our results show that two dipolarly coupled spin-torque vortex oscillators follow ideal oscillator systems described by Eq. (\ref{eq1}), a pre-assumption for studies based on the Kuramoto model \cite{FlovikSREP2016}. We confirm that the dipolar coupling strength can be tuned by a factor greater than two with bistable polarity states \cite{AbreuAraujoPRB2015,locatelliSREP2015}, providing a unique freedom to manipulate the collective dynamics. For instance, a new propagating wave mode has been predicted in oscillator arrays with both attracting and repulsive interactions \cite{HongPRL2011}, which can be realized with the two different polarity alignments. In addition we learn about the nonlinearities in spin-torque nano-oscillators. Finite phase shifts $\gamma_i^{NL}$ are measured, as predicted in theory \cite{ZhouJAP2007,SlavinIEEE2009} and identified in similar systems \cite{LebrunPRL2015,LebrunArXiv2016}. This indicates that practical oscillator networks fall into the Sakaguchi-Kuramoto regime (Eq. \ref{eq0} with nonzero $\beta_i$), in which synchronization can be destroyed by the phase detunings at medium $\Omega$ \cite{OlehPRL2012}.

In summary, we have developed a novel approach to study coupled oscillators with an external ac drive. By controlling the relative phases between the ac source and one phase-locked oscillator, we acquire not only the strength but also the phase information of the inter-oscillator coupling. This probing technique is not restricted to spin-torque oscillators and microwave field, but applicable to all coupled oscillator systems and ac drives. By extending their understanding, it is also useful for further manipulation and investigation of collective dynamics in large arrays of auto-oscillators.

We thank R. Lebrun, V. V. Naletov and A. N. Slavin for fruitful discussions. We acknowledge the MEMOS project ANR-14-CE26-0021 for financial support.


\begin{thebibliography}{56}%
\makeatletter
\providecommand \@ifxundefined [1]{%
 \@ifx{#1\undefined}
}%
\providecommand \@ifnum [1]{%
 \ifnum #1\expandafter \@firstoftwo
 \else \expandafter \@secondoftwo
 \fi
}%
\providecommand \@ifx [1]{%
 \ifx #1\expandafter \@firstoftwo
 \else \expandafter \@secondoftwo
 \fi
}%
\providecommand \natexlab [1]{#1}%
\providecommand \enquote  [1]{``#1''}%
\providecommand \bibnamefont  [1]{#1}%
\providecommand \bibfnamefont [1]{#1}%
\providecommand \citenamefont [1]{#1}%
\providecommand \href@noop [0]{\@secondoftwo}%
\providecommand \href [0]{\begingroup \@sanitize@url \@href}%
\providecommand \@href[1]{\@@startlink{#1}\@@href}%
\providecommand \@@href[1]{\endgroup#1\@@endlink}%
\providecommand \@sanitize@url [0]{\catcode `\\12\catcode `\$12\catcode
  `\&12\catcode `\#12\catcode `\^12\catcode `\_12\catcode `\%12\relax}%
\providecommand \@@startlink[1]{}%
\providecommand \@@endlink[0]{}%
\providecommand \url  [0]{\begingroup\@sanitize@url \@url }%
\providecommand \@url [1]{\endgroup\@href {#1}{\urlprefix }}%
\providecommand \urlprefix  [0]{URL }%
\providecommand \Eprint [0]{\href }%
\providecommand \doibase [0]{http://dx.doi.org/}%
\providecommand \selectlanguage [0]{\@gobble}%
\providecommand \bibinfo  [0]{\@secondoftwo}%
\providecommand \bibfield  [0]{\@secondoftwo}%
\providecommand \translation [1]{[#1]}%
\providecommand \BibitemOpen [0]{}%
\providecommand \bibitemStop [0]{}%
\providecommand \bibitemNoStop [0]{.\EOS\space}%
\providecommand \EOS [0]{\spacefactor3000\relax}%
\providecommand \BibitemShut  [1]{\csname bibitem#1\endcsname}%
\let\auto@bib@innerbib\@empty
\bibitem [{\citenamefont {Pikovsky}\ \emph {et~al.}(2001)\citenamefont
  {Pikovsky}, \citenamefont {Rosenblum},\ and\ \citenamefont
  {Kurths}}]{PikovskyBook2001}%
  \BibitemOpen
  \bibfield  {author} {\bibinfo {author} {\bibfnamefont {A.}~\bibnamefont
  {Pikovsky}}, \bibinfo {author} {\bibfnamefont {M.}~\bibnamefont {Rosenblum}},
  \ and\ \bibinfo {author} {\bibfnamefont {J.}~\bibnamefont {Kurths}},\
  }\href@noop {} {\emph {\bibinfo {title} {Synchronization: A universal concept
  in nonlinear sciences}}}\ (\bibinfo  {publisher} {Cambridge University Press,
  Cambridge, UK},\ \bibinfo {year} {2001})\BibitemShut {NoStop}%
\bibitem [{\citenamefont {Wiesenfeld}\ \emph {et~al.}(1996)\citenamefont
  {Wiesenfeld}, \citenamefont {Colet},\ and\ \citenamefont
  {Strogatz}}]{WiesenfeldPRL1996}%
  \BibitemOpen
  \bibfield  {author} {\bibinfo {author} {\bibfnamefont {K.}~\bibnamefont
  {Wiesenfeld}}, \bibinfo {author} {\bibfnamefont {P.}~\bibnamefont {Colet}}, \
  and\ \bibinfo {author} {\bibfnamefont {S.~H.}\ \bibnamefont {Strogatz}},\
  }\href@noop {} {\bibfield  {journal} {\bibinfo  {journal} {Phys. Rev. Lett.}\
  }\textbf {\bibinfo {volume} {76}},\ \bibinfo {pages} {404} (\bibinfo {year}
  {1996})}\BibitemShut {NoStop}%
\bibitem [{\citenamefont {Shim}\ \emph {et~al.}(2007)\citenamefont {Shim},
  \citenamefont {Imboden},\ and\ \citenamefont {Mohanty}}]{ShimScience2007}%
  \BibitemOpen
  \bibfield  {author} {\bibinfo {author} {\bibfnamefont {S.-B.}\ \bibnamefont
  {Shim}}, \bibinfo {author} {\bibfnamefont {M.}~\bibnamefont {Imboden}}, \
  and\ \bibinfo {author} {\bibfnamefont {P.}~\bibnamefont {Mohanty}},\
  }\href@noop {} {\bibfield  {journal} {\bibinfo  {journal} {Science}\ }\textbf
  {\bibinfo {volume} {316}},\ \bibinfo {pages} {95} (\bibinfo {year}
  {2007})}\BibitemShut {NoStop}%
\bibitem [{\citenamefont {Heinrich}\ \emph {et~al.}(2011)\citenamefont
  {Heinrich}, \citenamefont {Ludwig}, \citenamefont {Qian}, \citenamefont
  {Kubala},\ and\ \citenamefont {Marquardt}}]{HeinrichPRL2011}%
  \BibitemOpen
  \bibfield  {author} {\bibinfo {author} {\bibfnamefont {G.}~\bibnamefont
  {Heinrich}}, \bibinfo {author} {\bibfnamefont {M.}~\bibnamefont {Ludwig}},
  \bibinfo {author} {\bibfnamefont {J.}~\bibnamefont {Qian}}, \bibinfo {author}
  {\bibfnamefont {B.}~\bibnamefont {Kubala}}, \ and\ \bibinfo {author}
  {\bibfnamefont {F.}~\bibnamefont {Marquardt}},\ }\href@noop {} {\bibfield
  {journal} {\bibinfo  {journal} {Phys. Rev. Lett.}\ }\textbf {\bibinfo
  {volume} {107}},\ \bibinfo {pages} {043603} (\bibinfo {year}
  {2011})}\BibitemShut {NoStop}%
\bibitem [{\citenamefont {Zhang}\ \emph {et~al.}(2012)\citenamefont {Zhang},
  \citenamefont {Wiederhecker}, \citenamefont {Manipatruni}, \citenamefont
  {Barnard}, \citenamefont {McEuen},\ and\ \citenamefont
  {Lipson}}]{MZhangPRL2012}%
  \BibitemOpen
  \bibfield  {author} {\bibinfo {author} {\bibfnamefont {M.}~\bibnamefont
  {Zhang}}, \bibinfo {author} {\bibfnamefont {G.~S.}\ \bibnamefont
  {Wiederhecker}}, \bibinfo {author} {\bibfnamefont {S.}~\bibnamefont
  {Manipatruni}}, \bibinfo {author} {\bibfnamefont {A.}~\bibnamefont
  {Barnard}}, \bibinfo {author} {\bibfnamefont {P.}~\bibnamefont {McEuen}}, \
  and\ \bibinfo {author} {\bibfnamefont {M.}~\bibnamefont {Lipson}},\
  }\href@noop {} {\bibfield  {journal} {\bibinfo  {journal} {Phys. Rev. Lett.}\
  }\textbf {\bibinfo {volume} {109}},\ \bibinfo {pages} {233906} (\bibinfo
  {year} {2012})}\BibitemShut {NoStop}%
\bibitem [{\citenamefont {Kaka}\ \emph {et~al.}(2005)\citenamefont {Kaka},
  \citenamefont {Pufall}, \citenamefont {Rippard}, \citenamefont {Silva},
  \citenamefont {Russek},\ and\ \citenamefont {Katine}}]{KakaNature2005}%
  \BibitemOpen
  \bibfield  {author} {\bibinfo {author} {\bibfnamefont {S.}~\bibnamefont
  {Kaka}}, \bibinfo {author} {\bibfnamefont {M.~R.}\ \bibnamefont {Pufall}},
  \bibinfo {author} {\bibfnamefont {W.~H.}\ \bibnamefont {Rippard}}, \bibinfo
  {author} {\bibfnamefont {T.~J.}\ \bibnamefont {Silva}}, \bibinfo {author}
  {\bibfnamefont {S.~E.}\ \bibnamefont {Russek}}, \ and\ \bibinfo {author}
  {\bibfnamefont {J.~A.}\ \bibnamefont {Katine}},\ }\href@noop {} {\bibfield
  {journal} {\bibinfo  {journal} {Nature}\ }\textbf {\bibinfo {volume} {437}},\
  \bibinfo {pages} {389} (\bibinfo {year} {2005})}\BibitemShut {NoStop}%
\bibitem [{\citenamefont {Mancoff}\ \emph {et~al.}(2005)\citenamefont
  {Mancoff}, \citenamefont {Rizzo}, \citenamefont {Engel},\ and\ \citenamefont
  {Tehrani}}]{MancoffNature2005}%
  \BibitemOpen
  \bibfield  {author} {\bibinfo {author} {\bibfnamefont {F.~B.}\ \bibnamefont
  {Mancoff}}, \bibinfo {author} {\bibfnamefont {N.~D.}\ \bibnamefont {Rizzo}},
  \bibinfo {author} {\bibfnamefont {B.~N.}\ \bibnamefont {Engel}}, \ and\
  \bibinfo {author} {\bibfnamefont {S.}~\bibnamefont {Tehrani}},\ }\href@noop
  {} {\bibfield  {journal} {\bibinfo  {journal} {Nature}\ }\textbf {\bibinfo
  {volume} {437}},\ \bibinfo {pages} {393} (\bibinfo {year}
  {2005})}\BibitemShut {NoStop}%
\bibitem [{\citenamefont {Kiss}\ \emph {et~al.}(2002)\citenamefont {Kiss},
  \citenamefont {Zhai},\ and\ \citenamefont {Hudson}}]{KissScience2002}%
  \BibitemOpen
  \bibfield  {author} {\bibinfo {author} {\bibfnamefont {I.~Z.}\ \bibnamefont
  {Kiss}}, \bibinfo {author} {\bibfnamefont {Y.}~\bibnamefont {Zhai}}, \ and\
  \bibinfo {author} {\bibfnamefont {J.~L.}\ \bibnamefont {Hudson}},\
  }\href@noop {} {\bibfield  {journal} {\bibinfo  {journal} {Science}\ }\textbf
  {\bibinfo {volume} {296}},\ \bibinfo {pages} {1676} (\bibinfo {year}
  {2002})}\BibitemShut {NoStop}%
\bibitem [{\citenamefont {Tinsley}\ \emph {et~al.}(2012)\citenamefont
  {Tinsley}, \citenamefont {Nkomo},\ and\ \citenamefont
  {Showalter}}]{TinsleyNphys2012}%
  \BibitemOpen
  \bibfield  {author} {\bibinfo {author} {\bibfnamefont {M.~R.}\ \bibnamefont
  {Tinsley}}, \bibinfo {author} {\bibfnamefont {S.}~\bibnamefont {Nkomo}}, \
  and\ \bibinfo {author} {\bibfnamefont {K.}~\bibnamefont {Showalter}},\
  }\href@noop {} {\bibfield  {journal} {\bibinfo  {journal} {Nat. Phys.}\
  }\textbf {\bibinfo {volume} {8}},\ \bibinfo {pages} {662} (\bibinfo {year}
  {2012})}\BibitemShut {NoStop}%
\bibitem [{\citenamefont {Hartwell}\ \emph {et~al.}(1999)\citenamefont
  {Hartwell}, \citenamefont {Hopfield}, \citenamefont {Leibler},\ and\
  \citenamefont {Murray}}]{HartwellNature1999}%
  \BibitemOpen
  \bibfield  {author} {\bibinfo {author} {\bibfnamefont {L.~H.}\ \bibnamefont
  {Hartwell}}, \bibinfo {author} {\bibfnamefont {J.~J.}\ \bibnamefont
  {Hopfield}}, \bibinfo {author} {\bibfnamefont {S.}~\bibnamefont {Leibler}}, \
  and\ \bibinfo {author} {\bibfnamefont {A.~W.}\ \bibnamefont {Murray}},\
  }\href@noop {} {\bibfield  {journal} {\bibinfo  {journal} {Nature}\ }\textbf
  {\bibinfo {volume} {402}},\ \bibinfo {pages} {C47} (\bibinfo {year}
  {1999})}\BibitemShut {NoStop}%
\bibitem [{\citenamefont {Locatelli}\ \emph {et~al.}(2013)\citenamefont
  {Locatelli}, \citenamefont {Cros},\ and\ \citenamefont
  {Grollier}}]{LocatelliNmat2013}%
  \BibitemOpen
  \bibfield  {author} {\bibinfo {author} {\bibfnamefont {N.}~\bibnamefont
  {Locatelli}}, \bibinfo {author} {\bibfnamefont {V.}~\bibnamefont {Cros}}, \
  and\ \bibinfo {author} {\bibfnamefont {J.}~\bibnamefont {Grollier}},\
  }\href@noop {} {\bibfield  {journal} {\bibinfo  {journal} {Nat. Mater.}\
  }\textbf {\bibinfo {volume} {13}},\ \bibinfo {pages} {11} (\bibinfo {year}
  {2013})}\BibitemShut {NoStop}%
\bibitem [{\citenamefont {Grollier}\ \emph {et~al.}(2016)\citenamefont
  {Grollier}, \citenamefont {Querlioz},\ and\ \citenamefont
  {Stiles}}]{GrollierProcIEEE2016}%
  \BibitemOpen
  \bibfield  {author} {\bibinfo {author} {\bibfnamefont {J.}~\bibnamefont
  {Grollier}}, \bibinfo {author} {\bibfnamefont {D.}~\bibnamefont {Querlioz}},
  \ and\ \bibinfo {author} {\bibfnamefont {M.~D.}\ \bibnamefont {Stiles}},\
  }\href@noop {} {\bibfield  {journal} {\bibinfo  {journal} {Proc. IEEE}\
  }\textbf {\bibinfo {volume} {104}},\ \bibinfo {pages} {2024} (\bibinfo {year}
  {2016})}\BibitemShut {NoStop}%
\bibitem [{\citenamefont {Kuramoto}(1984)}]{KuramotoBook1984}%
  \BibitemOpen
  \bibfield  {author} {\bibinfo {author} {\bibfnamefont {Y.}~\bibnamefont
  {Kuramoto}},\ }\href@noop {} {\emph {\bibinfo {title} {Chemical Oscillations,
  Waves, and Turbulence}}}\ (\bibinfo  {publisher} {Springer, Berlin},\
  \bibinfo {year} {1984})\BibitemShut {NoStop}%
\bibitem [{\citenamefont {Acebr\'{o}n}\ \emph {et~al.}(2005)\citenamefont
  {Acebr\'{o}n}, \citenamefont {Bonilla}, \citenamefont {P\'{e}rez-Vicente},
  \citenamefont {Ritort},\ and\ \citenamefont {Spigler}}]{AcebronRMP2005}%
  \BibitemOpen
  \bibfield  {author} {\bibinfo {author} {\bibfnamefont {J.~A.}\ \bibnamefont
  {Acebr\'{o}n}}, \bibinfo {author} {\bibfnamefont {L.~L.}\ \bibnamefont
  {Bonilla}}, \bibinfo {author} {\bibfnamefont {C.~J.}\ \bibnamefont
  {P\'{e}rez-Vicente}}, \bibinfo {author} {\bibfnamefont {F.}~\bibnamefont
  {Ritort}}, \ and\ \bibinfo {author} {\bibfnamefont {R.}~\bibnamefont
  {Spigler}},\ }\href@noop {} {\bibfield  {journal} {\bibinfo  {journal} {Rev.
  Mod. Phys.}\ }\textbf {\bibinfo {volume} {77}},\ \bibinfo {pages} {137}
  (\bibinfo {year} {2005})}\BibitemShut {NoStop}%
\bibitem [{\citenamefont {Slavin}\ and\ \citenamefont
  {Tiberkevich}(2009)}]{SlavinIEEE2009}%
  \BibitemOpen
  \bibfield  {author} {\bibinfo {author} {\bibfnamefont {A.}~\bibnamefont
  {Slavin}}\ and\ \bibinfo {author} {\bibfnamefont {V.}~\bibnamefont
  {Tiberkevich}},\ }\href@noop {} {\bibfield  {journal} {\bibinfo  {journal}
  {IEEE Trans. Magn.}\ }\textbf {\bibinfo {volume} {45}},\ \bibinfo {pages}
  {1875} (\bibinfo {year} {2009})}\BibitemShut {NoStop}%
\bibitem [{\citenamefont {Ruotolo}\ \emph {et~al.}(2009)\citenamefont
  {Ruotolo}, \citenamefont {Cros}, \citenamefont {Georges}, \citenamefont
  {Dussaux}, \citenamefont {Grollier}, \citenamefont {Deranlot}, \citenamefont
  {Guillemet}, \citenamefont {Bouzehouane}, \citenamefont {Fusil},\ and\
  \citenamefont {Fert}}]{RuotoloNnano2009}%
  \BibitemOpen
  \bibfield  {author} {\bibinfo {author} {\bibfnamefont {A.}~\bibnamefont
  {Ruotolo}}, \bibinfo {author} {\bibfnamefont {V.}~\bibnamefont {Cros}},
  \bibinfo {author} {\bibfnamefont {B.}~\bibnamefont {Georges}}, \bibinfo
  {author} {\bibfnamefont {A.}~\bibnamefont {Dussaux}}, \bibinfo {author}
  {\bibfnamefont {J.}~\bibnamefont {Grollier}}, \bibinfo {author}
  {\bibfnamefont {C.}~\bibnamefont {Deranlot}}, \bibinfo {author}
  {\bibfnamefont {R.}~\bibnamefont {Guillemet}}, \bibinfo {author}
  {\bibfnamefont {K.}~\bibnamefont {Bouzehouane}}, \bibinfo {author}
  {\bibfnamefont {S.}~\bibnamefont {Fusil}}, \ and\ \bibinfo {author}
  {\bibfnamefont {A.}~\bibnamefont {Fert}},\ }\href@noop {} {\bibfield
  {journal} {\bibinfo  {journal} {Nature Nano.}\ }\textbf {\bibinfo {volume}
  {4}},\ \bibinfo {pages} {528} (\bibinfo {year} {2009})}\BibitemShut {NoStop}%
\bibitem [{\citenamefont {Sani}\ \emph {et~al.}(2013)\citenamefont {Sani},
  \citenamefont {Persson}, \citenamefont {Mohseni}, \citenamefont {Pogoryelov},
  \citenamefont {Muduli}, \citenamefont {Eklund}, \citenamefont {Malm},
  \citenamefont {K\"{a}ll}, \citenamefont {Dmitriev},\ and\ \citenamefont
  {\r{A}kerman}}]{SaniNcomm2013}%
  \BibitemOpen
  \bibfield  {author} {\bibinfo {author} {\bibfnamefont {S.}~\bibnamefont
  {Sani}}, \bibinfo {author} {\bibfnamefont {J.}~\bibnamefont {Persson}},
  \bibinfo {author} {\bibfnamefont {S.~M.}\ \bibnamefont {Mohseni}}, \bibinfo
  {author} {\bibfnamefont {Y.}~\bibnamefont {Pogoryelov}}, \bibinfo {author}
  {\bibfnamefont {P.~K.}\ \bibnamefont {Muduli}}, \bibinfo {author}
  {\bibfnamefont {A.}~\bibnamefont {Eklund}}, \bibinfo {author} {\bibfnamefont
  {G.}~\bibnamefont {Malm}}, \bibinfo {author} {\bibfnamefont {M.}~\bibnamefont
  {K\"{a}ll}}, \bibinfo {author} {\bibfnamefont {A.}~\bibnamefont {Dmitriev}},
  \ and\ \bibinfo {author} {\bibfnamefont {J.}~\bibnamefont {\r{A}kerman}},\
  }\href@noop {} {\bibfield  {journal} {\bibinfo  {journal} {Nature Comm.}\
  }\textbf {\bibinfo {volume} {4}},\ \bibinfo {pages} {2731} (\bibinfo {year}
  {2013})}\BibitemShut {NoStop}%
\bibitem [{\citenamefont {Locatelli}\ \emph {et~al.}(2015)\citenamefont
  {Locatelli}, \citenamefont {Hamadeh}, \citenamefont {Abreu~Araujo},
  \citenamefont {Belanovsky}, \citenamefont {Skirdkov}, \citenamefont {Lebrun},
  \citenamefont {Naletov}, \citenamefont {Zvezdin}, \citenamefont {Mu\~{n}oz},
  \citenamefont {Grollier}, \citenamefont {Klein}, \citenamefont {Cros},\ and\
  \citenamefont {de~Loubens}}]{locatelliSREP2015}%
  \BibitemOpen
  \bibfield  {author} {\bibinfo {author} {\bibfnamefont {N.}~\bibnamefont
  {Locatelli}}, \bibinfo {author} {\bibfnamefont {A.}~\bibnamefont {Hamadeh}},
  \bibinfo {author} {\bibfnamefont {F.}~\bibnamefont {Abreu~Araujo}}, \bibinfo
  {author} {\bibfnamefont {A.~D.}\ \bibnamefont {Belanovsky}}, \bibinfo
  {author} {\bibfnamefont {P.~N.}\ \bibnamefont {Skirdkov}}, \bibinfo {author}
  {\bibfnamefont {R.}~\bibnamefont {Lebrun}}, \bibinfo {author} {\bibfnamefont
  {V.~V.}\ \bibnamefont {Naletov}}, \bibinfo {author} {\bibfnamefont {K.~A.}\
  \bibnamefont {Zvezdin}}, \bibinfo {author} {\bibfnamefont {M.}~\bibnamefont
  {Mu\~{n}oz}}, \bibinfo {author} {\bibfnamefont {J.}~\bibnamefont {Grollier}},
  \bibinfo {author} {\bibfnamefont {O.}~\bibnamefont {Klein}}, \bibinfo
  {author} {\bibfnamefont {V.}~\bibnamefont {Cros}}, \ and\ \bibinfo {author}
  {\bibfnamefont {G.}~\bibnamefont {de~Loubens}},\ }\href@noop {} {\bibfield
  {journal} {\bibinfo  {journal} {Sci. Rep.}\ }\textbf {\bibinfo {volume}
  {5}},\ \bibinfo {pages} {17039} (\bibinfo {year} {2015})}\BibitemShut
  {NoStop}%
\bibitem [{\citenamefont {Lebrun}\ \emph {et~al.}(2016)\citenamefont {Lebrun},
  \citenamefont {Tsunegi}, \citenamefont {Bortolotti}, \citenamefont {Kubota},
  \citenamefont {Jenkins}, \citenamefont {Romera}, \citenamefont {Yakushiji},
  \citenamefont {Fukushima}, \citenamefont {Grollier}, \citenamefont {Yuasa},\
  and\ \citenamefont {Cros}}]{LebrunArXiv2016}%
  \BibitemOpen
  \bibfield  {author} {\bibinfo {author} {\bibfnamefont {R.}~\bibnamefont
  {Lebrun}}, \bibinfo {author} {\bibfnamefont {S.}~\bibnamefont {Tsunegi}},
  \bibinfo {author} {\bibfnamefont {P.}~\bibnamefont {Bortolotti}}, \bibinfo
  {author} {\bibfnamefont {H.}~\bibnamefont {Kubota}}, \bibinfo {author}
  {\bibfnamefont {A.~S.}\ \bibnamefont {Jenkins}}, \bibinfo {author}
  {\bibfnamefont {M.}~\bibnamefont {Romera}}, \bibinfo {author} {\bibfnamefont
  {K.}~\bibnamefont {Yakushiji}}, \bibinfo {author} {\bibfnamefont
  {A.}~\bibnamefont {Fukushima}}, \bibinfo {author} {\bibfnamefont
  {J.}~\bibnamefont {Grollier}}, \bibinfo {author} {\bibfnamefont
  {S.}~\bibnamefont {Yuasa}}, \ and\ \bibinfo {author} {\bibfnamefont
  {V.}~\bibnamefont {Cros}},\ }\href@noop {} {\bibfield  {journal} {\bibinfo
  {journal} {arXiv}\ ,\ \bibinfo {pages} {1601.01247}} (\bibinfo {year}
  {2016})}\BibitemShut {NoStop}%
\bibitem [{\citenamefont {Tiberkevich}\ \emph {et~al.}(2009)\citenamefont
  {Tiberkevich}, \citenamefont {Slavin}, \citenamefont {Bankowski},\ and\
  \citenamefont {Gerhart}}]{TiberkevichAPL2009}%
  \BibitemOpen
  \bibfield  {author} {\bibinfo {author} {\bibfnamefont {V.}~\bibnamefont
  {Tiberkevich}}, \bibinfo {author} {\bibfnamefont {A.}~\bibnamefont {Slavin}},
  \bibinfo {author} {\bibfnamefont {E.}~\bibnamefont {Bankowski}}, \ and\
  \bibinfo {author} {\bibfnamefont {G.}~\bibnamefont {Gerhart}},\ }\href@noop
  {} {\bibfield  {journal} {\bibinfo  {journal} {Appl. Phys. Lett.}\ }\textbf
  {\bibinfo {volume} {95}},\ \bibinfo {pages} {262505} (\bibinfo {year}
  {2009})}\BibitemShut {NoStop}%
\bibitem [{\citenamefont {Omel'chenko}\ and\ \citenamefont
  {Wolfrum}(2012)}]{OlehPRL2012}%
  \BibitemOpen
  \bibfield  {author} {\bibinfo {author} {\bibfnamefont {O.~E.}\ \bibnamefont
  {Omel'chenko}}\ and\ \bibinfo {author} {\bibfnamefont {M.}~\bibnamefont
  {Wolfrum}},\ }\href {\doibase 10.1103/PhysRevLett.109.164101} {\bibfield
  {journal} {\bibinfo  {journal} {Phys. Rev. Lett.}\ }\textbf {\bibinfo
  {volume} {109}},\ \bibinfo {pages} {164101} (\bibinfo {year}
  {2012})}\BibitemShut {NoStop}%
\bibitem [{\citenamefont {Csaba}\ \emph {et~al.}(2012)\citenamefont {Csaba},
  \citenamefont {Pufall}, \citenamefont {Nikonov}, \citenamefont {Bourianoff},
  \citenamefont {Horvath}, \citenamefont {Roska},\ and\ \citenamefont
  {Porod}}]{csaba12}%
  \BibitemOpen
  \bibfield  {author} {\bibinfo {author} {\bibfnamefont {G.}~\bibnamefont
  {Csaba}}, \bibinfo {author} {\bibfnamefont {M.}~\bibnamefont {Pufall}},
  \bibinfo {author} {\bibfnamefont {D.}~\bibnamefont {Nikonov}}, \bibinfo
  {author} {\bibfnamefont {G.}~\bibnamefont {Bourianoff}}, \bibinfo {author}
  {\bibfnamefont {A.}~\bibnamefont {Horvath}}, \bibinfo {author} {\bibfnamefont
  {T.}~\bibnamefont {Roska}}, \ and\ \bibinfo {author} {\bibfnamefont
  {W.}~\bibnamefont {Porod}},\ }in\ \href {\doibase 10.1109/CNNA.2012.6331474}
  {\emph {\bibinfo {booktitle} {Cellular Nanoscale Networks and Their
  Applications (CNNA), 2012 13th International Workshop on}}}\ (\bibinfo {year}
  {2012})\ pp.\ \bibinfo {pages} {1--2}\BibitemShut {NoStop}%
\bibitem [{\citenamefont {Nikonov}\ \emph {et~al.}(2015)\citenamefont
  {Nikonov}, \citenamefont {Csaba}, \citenamefont {Porod}, \citenamefont
  {Shibata}, \citenamefont {Voils}, \citenamefont {Hammerstrom}, \citenamefont
  {Young},\ and\ \citenamefont {Bourianoff}}]{nikonov15}%
  \BibitemOpen
  \bibfield  {author} {\bibinfo {author} {\bibfnamefont {D.~E.}\ \bibnamefont
  {Nikonov}}, \bibinfo {author} {\bibfnamefont {G.}~\bibnamefont {Csaba}},
  \bibinfo {author} {\bibfnamefont {W.}~\bibnamefont {Porod}}, \bibinfo
  {author} {\bibfnamefont {T.}~\bibnamefont {Shibata}}, \bibinfo {author}
  {\bibfnamefont {D.}~\bibnamefont {Voils}}, \bibinfo {author} {\bibfnamefont
  {D.}~\bibnamefont {Hammerstrom}}, \bibinfo {author} {\bibfnamefont {I.~A.}\
  \bibnamefont {Young}}, \ and\ \bibinfo {author} {\bibfnamefont {G.~I.}\
  \bibnamefont {Bourianoff}},\ }\href {\doibase 10.1109/JXCDC.2015.2504049}
  {\bibfield  {journal} {\bibinfo  {journal} {IEEE Journal on Exploratory
  Solid-State Computational Devices and Circuits}\ }\textbf {\bibinfo {volume}
  {1}},\ \bibinfo {pages} {85} (\bibinfo {year} {2015})}\BibitemShut {NoStop}%
\bibitem [{\citenamefont {Kiselev}\ \emph {et~al.}(2003)\citenamefont
  {Kiselev}, \citenamefont {Sankey}, \citenamefont {Krivorotov}, \citenamefont
  {Emley}, \citenamefont {Schoelkopf}, \citenamefont {Buhrman},\ and\
  \citenamefont {Ralph}}]{KiselevNature2003}%
  \BibitemOpen
  \bibfield  {author} {\bibinfo {author} {\bibfnamefont {S.~I.}\ \bibnamefont
  {Kiselev}}, \bibinfo {author} {\bibfnamefont {J.~C.}\ \bibnamefont {Sankey}},
  \bibinfo {author} {\bibfnamefont {I.~N.}\ \bibnamefont {Krivorotov}},
  \bibinfo {author} {\bibfnamefont {N.~C.}\ \bibnamefont {Emley}}, \bibinfo
  {author} {\bibfnamefont {R.~J.}\ \bibnamefont {Schoelkopf}}, \bibinfo
  {author} {\bibfnamefont {R.~A.}\ \bibnamefont {Buhrman}}, \ and\ \bibinfo
  {author} {\bibfnamefont {D.~C.}\ \bibnamefont {Ralph}},\ }\href@noop {}
  {\bibfield  {journal} {\bibinfo  {journal} {Nature}\ }\textbf {\bibinfo
  {volume} {425}},\ \bibinfo {pages} {380} (\bibinfo {year}
  {2003})}\BibitemShut {NoStop}%
\bibitem [{\citenamefont {Rippard}\ \emph {et~al.}(2005)\citenamefont
  {Rippard}, \citenamefont {Pufall}, \citenamefont {Kaka}, \citenamefont
  {Silva}, \citenamefont {Russek},\ and\ \citenamefont
  {Katine}}]{RippardPRL2005}%
  \BibitemOpen
  \bibfield  {author} {\bibinfo {author} {\bibfnamefont {W.~H.}\ \bibnamefont
  {Rippard}}, \bibinfo {author} {\bibfnamefont {M.~R.}\ \bibnamefont {Pufall}},
  \bibinfo {author} {\bibfnamefont {S.}~\bibnamefont {Kaka}}, \bibinfo {author}
  {\bibfnamefont {T.~J.}\ \bibnamefont {Silva}}, \bibinfo {author}
  {\bibfnamefont {S.~E.}\ \bibnamefont {Russek}}, \ and\ \bibinfo {author}
  {\bibfnamefont {J.~A.}\ \bibnamefont {Katine}},\ }\href@noop {} {\bibfield
  {journal} {\bibinfo  {journal} {Phys. Rev. Lett.}\ }\textbf {\bibinfo
  {volume} {95}},\ \bibinfo {pages} {067203} (\bibinfo {year}
  {2005})}\BibitemShut {NoStop}%
\bibitem [{\citenamefont {Georges}\ \emph {et~al.}(2008)\citenamefont
  {Georges}, \citenamefont {Grollier}, \citenamefont {Darques}, \citenamefont
  {Cros}, \citenamefont {Deranlot}, \citenamefont {Marcilhac}, \citenamefont
  {Faini},\ and\ \citenamefont {Fert}}]{GeorgesPRL2008}%
  \BibitemOpen
  \bibfield  {author} {\bibinfo {author} {\bibfnamefont {B.}~\bibnamefont
  {Georges}}, \bibinfo {author} {\bibfnamefont {J.}~\bibnamefont {Grollier}},
  \bibinfo {author} {\bibfnamefont {M.}~\bibnamefont {Darques}}, \bibinfo
  {author} {\bibfnamefont {V.}~\bibnamefont {Cros}}, \bibinfo {author}
  {\bibfnamefont {C.}~\bibnamefont {Deranlot}}, \bibinfo {author}
  {\bibfnamefont {B.}~\bibnamefont {Marcilhac}}, \bibinfo {author}
  {\bibfnamefont {G.}~\bibnamefont {Faini}}, \ and\ \bibinfo {author}
  {\bibfnamefont {A.}~\bibnamefont {Fert}},\ }\href@noop {} {\bibfield
  {journal} {\bibinfo  {journal} {Phys. Rev. Lett.}\ }\textbf {\bibinfo
  {volume} {101}},\ \bibinfo {pages} {017201} (\bibinfo {year}
  {2008})}\BibitemShut {NoStop}%
\bibitem [{\citenamefont {Urazhdin}\ \emph {et~al.}(2010)\citenamefont
  {Urazhdin}, \citenamefont {Tabor}, \citenamefont {Tiberkevich},\ and\
  \citenamefont {Slavin}}]{UrazhdinPRL2010}%
  \BibitemOpen
  \bibfield  {author} {\bibinfo {author} {\bibfnamefont {S.}~\bibnamefont
  {Urazhdin}}, \bibinfo {author} {\bibfnamefont {P.}~\bibnamefont {Tabor}},
  \bibinfo {author} {\bibfnamefont {V.}~\bibnamefont {Tiberkevich}}, \ and\
  \bibinfo {author} {\bibfnamefont {A.}~\bibnamefont {Slavin}},\ }\href@noop {}
  {\bibfield  {journal} {\bibinfo  {journal} {Phys. Rev. Lett.}\ }\textbf
  {\bibinfo {volume} {105}},\ \bibinfo {pages} {104101} (\bibinfo {year}
  {2010})}\BibitemShut {NoStop}%
\bibitem [{\citenamefont {Quinsat}\ \emph {et~al.}(2011)\citenamefont
  {Quinsat}, \citenamefont {Sierra}, \citenamefont {Firastrau}, \citenamefont
  {Tiberkevich}, \citenamefont {Slavin}, \citenamefont {Gusakova},
  \citenamefont {Buda-Prejbeanu}, \citenamefont {Zarudniev}, \citenamefont
  {Michel}, \citenamefont {Ebels}, \citenamefont {Dieny}, \citenamefont
  {Cyrille}, \citenamefont {Katine}, \citenamefont {Mauri},\ and\ \citenamefont
  {Zeltser}}]{QuinsatAPL2011}%
  \BibitemOpen
  \bibfield  {author} {\bibinfo {author} {\bibfnamefont {M.}~\bibnamefont
  {Quinsat}}, \bibinfo {author} {\bibfnamefont {J.~F.}\ \bibnamefont {Sierra}},
  \bibinfo {author} {\bibfnamefont {I.}~\bibnamefont {Firastrau}}, \bibinfo
  {author} {\bibfnamefont {V.}~\bibnamefont {Tiberkevich}}, \bibinfo {author}
  {\bibfnamefont {A.}~\bibnamefont {Slavin}}, \bibinfo {author} {\bibfnamefont
  {D.}~\bibnamefont {Gusakova}}, \bibinfo {author} {\bibfnamefont {L.~D.}\
  \bibnamefont {Buda-Prejbeanu}}, \bibinfo {author} {\bibfnamefont
  {M.}~\bibnamefont {Zarudniev}}, \bibinfo {author} {\bibfnamefont {J.-P.}\
  \bibnamefont {Michel}}, \bibinfo {author} {\bibfnamefont {U.}~\bibnamefont
  {Ebels}}, \bibinfo {author} {\bibfnamefont {B.}~\bibnamefont {Dieny}},
  \bibinfo {author} {\bibfnamefont {M.-C.}\ \bibnamefont {Cyrille}}, \bibinfo
  {author} {\bibfnamefont {J.~A.}\ \bibnamefont {Katine}}, \bibinfo {author}
  {\bibfnamefont {D.}~\bibnamefont {Mauri}}, \ and\ \bibinfo {author}
  {\bibfnamefont {A.}~\bibnamefont {Zeltser}},\ }\href@noop {} {\bibfield
  {journal} {\bibinfo  {journal} {Appl. Phys. Lett.}\ }\textbf {\bibinfo
  {volume} {98}},\ \bibinfo {pages} {182503} (\bibinfo {year}
  {2011})}\BibitemShut {NoStop}%
\bibitem [{\citenamefont {Hamadeh}\ \emph
  {et~al.}(2014{\natexlab{a}})\citenamefont {Hamadeh}, \citenamefont
  {Locatelli}, \citenamefont {Naletov}, \citenamefont {Lebrun}, \citenamefont
  {de~Loubens}, \citenamefont {Grollier}, \citenamefont {Klein},\ and\
  \citenamefont {Cros}}]{HamadehAPL2014}%
  \BibitemOpen
  \bibfield  {author} {\bibinfo {author} {\bibfnamefont {A.}~\bibnamefont
  {Hamadeh}}, \bibinfo {author} {\bibfnamefont {N.}~\bibnamefont {Locatelli}},
  \bibinfo {author} {\bibfnamefont {V.~V.}\ \bibnamefont {Naletov}}, \bibinfo
  {author} {\bibfnamefont {R.}~\bibnamefont {Lebrun}}, \bibinfo {author}
  {\bibfnamefont {G.}~\bibnamefont {de~Loubens}}, \bibinfo {author}
  {\bibfnamefont {J.}~\bibnamefont {Grollier}}, \bibinfo {author}
  {\bibfnamefont {O.}~\bibnamefont {Klein}}, \ and\ \bibinfo {author}
  {\bibfnamefont {V.}~\bibnamefont {Cros}},\ }\href@noop {} {\bibfield
  {journal} {\bibinfo  {journal} {Appl. Phys. Lett.}\ }\textbf {\bibinfo
  {volume} {104}},\ \bibinfo {pages} {022408} (\bibinfo {year}
  {2014}{\natexlab{a}})}\BibitemShut {NoStop}%
\bibitem [{\citenamefont {Lebrun}\ \emph {et~al.}(2015)\citenamefont {Lebrun},
  \citenamefont {Jenkins}, \citenamefont {Dussaux}, \citenamefont {Locatelli},
  \citenamefont {Tsunegi}, \citenamefont {Grimaldi}, \citenamefont {Kubota},
  \citenamefont {Bortolotti}, \citenamefont {Yakushiji}, \citenamefont
  {Grollier}, \citenamefont {Fukushima}, \citenamefont {Yuasa},\ and\
  \citenamefont {Cros}}]{LebrunPRL2015}%
  \BibitemOpen
  \bibfield  {author} {\bibinfo {author} {\bibfnamefont {R.}~\bibnamefont
  {Lebrun}}, \bibinfo {author} {\bibfnamefont {A.}~\bibnamefont {Jenkins}},
  \bibinfo {author} {\bibfnamefont {A.}~\bibnamefont {Dussaux}}, \bibinfo
  {author} {\bibfnamefont {N.}~\bibnamefont {Locatelli}}, \bibinfo {author}
  {\bibfnamefont {S.}~\bibnamefont {Tsunegi}}, \bibinfo {author} {\bibfnamefont
  {E.}~\bibnamefont {Grimaldi}}, \bibinfo {author} {\bibfnamefont
  {H.}~\bibnamefont {Kubota}}, \bibinfo {author} {\bibfnamefont
  {P.}~\bibnamefont {Bortolotti}}, \bibinfo {author} {\bibfnamefont
  {K.}~\bibnamefont {Yakushiji}}, \bibinfo {author} {\bibfnamefont
  {J.}~\bibnamefont {Grollier}}, \bibinfo {author} {\bibfnamefont
  {A.}~\bibnamefont {Fukushima}}, \bibinfo {author} {\bibfnamefont
  {S.}~\bibnamefont {Yuasa}}, \ and\ \bibinfo {author} {\bibfnamefont
  {V.}~\bibnamefont {Cros}},\ }\href@noop {} {\bibfield  {journal} {\bibinfo
  {journal} {Phys. Rev. Lett.}\ }\textbf {\bibinfo {volume} {115}},\ \bibinfo
  {pages} {017201} (\bibinfo {year} {2015})}\BibitemShut {NoStop}%
\bibitem [{\citenamefont {Pribiag}\ \emph {et~al.}(2007)\citenamefont
  {Pribiag}, \citenamefont {Krivorotov}, \citenamefont {Fuchs}, \citenamefont
  {Braganca}, \citenamefont {Ozatay}, \citenamefont {Sankey}, \citenamefont
  {Ralph},\ and\ \citenamefont {Buhrman}}]{PribiagNphys2007}%
  \BibitemOpen
  \bibfield  {author} {\bibinfo {author} {\bibfnamefont {V.~S.}\ \bibnamefont
  {Pribiag}}, \bibinfo {author} {\bibfnamefont {I.~N.}\ \bibnamefont
  {Krivorotov}}, \bibinfo {author} {\bibfnamefont {G.~D.}\ \bibnamefont
  {Fuchs}}, \bibinfo {author} {\bibfnamefont {P.~M.}\ \bibnamefont {Braganca}},
  \bibinfo {author} {\bibfnamefont {O.}~\bibnamefont {Ozatay}}, \bibinfo
  {author} {\bibfnamefont {J.~C.}\ \bibnamefont {Sankey}}, \bibinfo {author}
  {\bibfnamefont {D.~C.}\ \bibnamefont {Ralph}}, \ and\ \bibinfo {author}
  {\bibfnamefont {R.~A.}\ \bibnamefont {Buhrman}},\ }\href@noop {} {\bibfield
  {journal} {\bibinfo  {journal} {Nature Phys.}\ }\textbf {\bibinfo {volume}
  {3}},\ \bibinfo {pages} {498} (\bibinfo {year} {2007})}\BibitemShut {NoStop}%
\bibitem [{\citenamefont {Dussaux}\ \emph {et~al.}(2010)\citenamefont
  {Dussaux}, \citenamefont {Georges}, \citenamefont {Grollier}, \citenamefont
  {Cros}, \citenamefont {Khvalkovskiy}, \citenamefont {Fukushima},
  \citenamefont {Konoto}, \citenamefont {Kubota}, \citenamefont {Yakushiji},
  \citenamefont {Yuasa}, \citenamefont {Zvezdin}, \citenamefont {Ando},\ and\
  \citenamefont {Fert}}]{DussauxNcomm2010}%
  \BibitemOpen
  \bibfield  {author} {\bibinfo {author} {\bibfnamefont {A.}~\bibnamefont
  {Dussaux}}, \bibinfo {author} {\bibfnamefont {B.}~\bibnamefont {Georges}},
  \bibinfo {author} {\bibfnamefont {J.}~\bibnamefont {Grollier}}, \bibinfo
  {author} {\bibfnamefont {V.}~\bibnamefont {Cros}}, \bibinfo {author}
  {\bibfnamefont {A.~V.}\ \bibnamefont {Khvalkovskiy}}, \bibinfo {author}
  {\bibfnamefont {A.}~\bibnamefont {Fukushima}}, \bibinfo {author}
  {\bibfnamefont {M.}~\bibnamefont {Konoto}}, \bibinfo {author} {\bibfnamefont
  {H.}~\bibnamefont {Kubota}}, \bibinfo {author} {\bibfnamefont
  {K.}~\bibnamefont {Yakushiji}}, \bibinfo {author} {\bibfnamefont
  {S.}~\bibnamefont {Yuasa}}, \bibinfo {author} {\bibfnamefont {K.~A.}\
  \bibnamefont {Zvezdin}}, \bibinfo {author} {\bibfnamefont {K.}~\bibnamefont
  {Ando}}, \ and\ \bibinfo {author} {\bibfnamefont {A.}~\bibnamefont {Fert}},\
  }\href@noop {} {\bibfield  {journal} {\bibinfo  {journal} {Nature Commun.}\
  }\textbf {\bibinfo {volume} {1}},\ \bibinfo {pages} {8} (\bibinfo {year}
  {2010})}\BibitemShut {NoStop}%
\bibitem [{\citenamefont {Locatelli}\ \emph {et~al.}(2011)\citenamefont
  {Locatelli}, \citenamefont {Naletov}, \citenamefont {Grollier}, \citenamefont
  {de~Loubens}, \citenamefont {Cros}, \citenamefont {Deranlot}, \citenamefont
  {Ulysse}, \citenamefont {Faini}, \citenamefont {Klein},\ and\ \citenamefont
  {Fert}}]{LocatelliAPL2011}%
  \BibitemOpen
  \bibfield  {author} {\bibinfo {author} {\bibfnamefont {N.}~\bibnamefont
  {Locatelli}}, \bibinfo {author} {\bibfnamefont {V.~V.}\ \bibnamefont
  {Naletov}}, \bibinfo {author} {\bibfnamefont {J.}~\bibnamefont {Grollier}},
  \bibinfo {author} {\bibfnamefont {G.}~\bibnamefont {de~Loubens}}, \bibinfo
  {author} {\bibfnamefont {V.}~\bibnamefont {Cros}}, \bibinfo {author}
  {\bibfnamefont {C.}~\bibnamefont {Deranlot}}, \bibinfo {author}
  {\bibfnamefont {C.}~\bibnamefont {Ulysse}}, \bibinfo {author} {\bibfnamefont
  {G.}~\bibnamefont {Faini}}, \bibinfo {author} {\bibfnamefont
  {O.}~\bibnamefont {Klein}}, \ and\ \bibinfo {author} {\bibfnamefont
  {A.}~\bibnamefont {Fert}},\ }\href@noop {} {\bibfield  {journal} {\bibinfo
  {journal} {Appl. Phys. Lett.}\ }\textbf {\bibinfo {volume} {98}},\ \bibinfo
  {pages} {062501} (\bibinfo {year} {2011})}\BibitemShut {NoStop}%
\bibitem [{\citenamefont {Hamadeh}\ \emph
  {et~al.}(2014{\natexlab{b}})\citenamefont {Hamadeh}, \citenamefont
  {Locatelli}, \citenamefont {Naletov}, \citenamefont {Lebrun}, \citenamefont
  {de~Loubens}, \citenamefont {Grollier}, \citenamefont {Klein},\ and\
  \citenamefont {Cros}}]{HamadehPRL2014}%
  \BibitemOpen
  \bibfield  {author} {\bibinfo {author} {\bibfnamefont {A.}~\bibnamefont
  {Hamadeh}}, \bibinfo {author} {\bibfnamefont {N.}~\bibnamefont {Locatelli}},
  \bibinfo {author} {\bibfnamefont {V.~V.}\ \bibnamefont {Naletov}}, \bibinfo
  {author} {\bibfnamefont {R.}~\bibnamefont {Lebrun}}, \bibinfo {author}
  {\bibfnamefont {G.}~\bibnamefont {de~Loubens}}, \bibinfo {author}
  {\bibfnamefont {J.}~\bibnamefont {Grollier}}, \bibinfo {author}
  {\bibfnamefont {O.}~\bibnamefont {Klein}}, \ and\ \bibinfo {author}
  {\bibfnamefont {V.}~\bibnamefont {Cros}},\ }\href {\doibase
  10.1103/PhysRevLett.112.257201} {\bibfield  {journal} {\bibinfo  {journal}
  {Phys. Rev. Lett.}\ }\textbf {\bibinfo {volume} {112}},\ \bibinfo {pages}
  {257201} (\bibinfo {year} {2014}{\natexlab{b}})}\BibitemShut {NoStop}%
\bibitem [{\citenamefont {Sluka}\ \emph {et~al.}(2015)\citenamefont {Sluka},
  \citenamefont {K\'{a}kay}, \citenamefont {Deac}, \citenamefont {B\"{u}rgler},
  \citenamefont {Schneider},\ and\ \citenamefont {Hertel}}]{SlukaNcomm2015}%
  \BibitemOpen
  \bibfield  {author} {\bibinfo {author} {\bibfnamefont {V.}~\bibnamefont
  {Sluka}}, \bibinfo {author} {\bibfnamefont {A.}~\bibnamefont {K\'{a}kay}},
  \bibinfo {author} {\bibfnamefont {A.~M.}\ \bibnamefont {Deac}}, \bibinfo
  {author} {\bibfnamefont {D.~E.}\ \bibnamefont {B\"{u}rgler}}, \bibinfo
  {author} {\bibfnamefont {C.~M.}\ \bibnamefont {Schneider}}, \ and\ \bibinfo
  {author} {\bibfnamefont {R.}~\bibnamefont {Hertel}},\ }\href@noop {}
  {\bibfield  {journal} {\bibinfo  {journal} {Nat. Comm.}\ }\textbf {\bibinfo
  {volume} {6}},\ \bibinfo {pages} {6409} (\bibinfo {year} {2015})}\BibitemShut
  {NoStop}%
\bibitem [{\citenamefont {de~Loubens}\ \emph {et~al.}(2009)\citenamefont
  {de~Loubens}, \citenamefont {Riegler}, \citenamefont {Pigeau}, \citenamefont
  {Lochner}, \citenamefont {Boust}, \citenamefont {Guslienko}, \citenamefont
  {Hurdequint}, \citenamefont {Molenkamp}, \citenamefont {Schmidt},
  \citenamefont {Slavin}, \citenamefont {Tiberkevich}, \citenamefont
  {Vukadinovic},\ and\ \citenamefont {Klein}}]{deLoubensPRL2009}%
  \BibitemOpen
  \bibfield  {author} {\bibinfo {author} {\bibfnamefont {G.}~\bibnamefont
  {de~Loubens}}, \bibinfo {author} {\bibfnamefont {A.}~\bibnamefont {Riegler}},
  \bibinfo {author} {\bibfnamefont {B.}~\bibnamefont {Pigeau}}, \bibinfo
  {author} {\bibfnamefont {F.}~\bibnamefont {Lochner}}, \bibinfo {author}
  {\bibfnamefont {F.}~\bibnamefont {Boust}}, \bibinfo {author} {\bibfnamefont
  {K.~Y.}\ \bibnamefont {Guslienko}}, \bibinfo {author} {\bibfnamefont
  {H.}~\bibnamefont {Hurdequint}}, \bibinfo {author} {\bibfnamefont {L.~W.}\
  \bibnamefont {Molenkamp}}, \bibinfo {author} {\bibfnamefont {G.}~\bibnamefont
  {Schmidt}}, \bibinfo {author} {\bibfnamefont {A.~N.}\ \bibnamefont {Slavin}},
  \bibinfo {author} {\bibfnamefont {V.~S.}\ \bibnamefont {Tiberkevich}},
  \bibinfo {author} {\bibfnamefont {N.}~\bibnamefont {Vukadinovic}}, \ and\
  \bibinfo {author} {\bibfnamefont {O.}~\bibnamefont {Klein}},\ }\href@noop {}
  {\bibfield  {journal} {\bibinfo  {journal} {Phys. Rev. Lett.}\ }\textbf
  {\bibinfo {volume} {102}},\ \bibinfo {pages} {177602} (\bibinfo {year}
  {2009})}\BibitemShut {NoStop}%
\bibitem [{\citenamefont {Shibata}\ \emph {et~al.}(2003)\citenamefont
  {Shibata}, \citenamefont {Shigeto},\ and\ \citenamefont
  {Otani}}]{ShibataPRB2003}%
  \BibitemOpen
  \bibfield  {author} {\bibinfo {author} {\bibfnamefont {J.}~\bibnamefont
  {Shibata}}, \bibinfo {author} {\bibfnamefont {K.}~\bibnamefont {Shigeto}}, \
  and\ \bibinfo {author} {\bibfnamefont {Y.}~\bibnamefont {Otani}},\
  }\href@noop {} {\bibfield  {journal} {\bibinfo  {journal} {Phys. Rev. B}\
  }\textbf {\bibinfo {volume} {67}},\ \bibinfo {pages} {224404} (\bibinfo
  {year} {2003})}\BibitemShut {NoStop}%
\bibitem [{\citenamefont {Sugimoto}\ \emph {et~al.}(2011)\citenamefont
  {Sugimoto}, \citenamefont {Fukuma}, \citenamefont {Kasai}, \citenamefont
  {Kimura}, \citenamefont {Barman},\ and\ \citenamefont
  {Otani}}]{SugimotoPRL2011}%
  \BibitemOpen
  \bibfield  {author} {\bibinfo {author} {\bibfnamefont {S.}~\bibnamefont
  {Sugimoto}}, \bibinfo {author} {\bibfnamefont {Y.}~\bibnamefont {Fukuma}},
  \bibinfo {author} {\bibfnamefont {S.}~\bibnamefont {Kasai}}, \bibinfo
  {author} {\bibfnamefont {T.}~\bibnamefont {Kimura}}, \bibinfo {author}
  {\bibfnamefont {A.}~\bibnamefont {Barman}}, \ and\ \bibinfo {author}
  {\bibfnamefont {Y.~C.}\ \bibnamefont {Otani}},\ }\href@noop {} {\bibfield
  {journal} {\bibinfo  {journal} {Phys. Rev. Lett.}\ }\textbf {\bibinfo
  {volume} {106}},\ \bibinfo {pages} {197203} (\bibinfo {year}
  {2011})}\BibitemShut {NoStop}%
\bibitem [{\citenamefont {Han}\ \emph {et~al.}(2013)\citenamefont {Han},
  \citenamefont {Vogel}, \citenamefont {Jung}, \citenamefont {Lee},
  \citenamefont {Weigand}, \citenamefont {Stoll}, \citenamefont {Sch\"{u}tz},
  \citenamefont {Fischer}, \citenamefont {Meier},\ and\ \citenamefont
  {Kim}}]{HanSREP2013}%
  \BibitemOpen
  \bibfield  {author} {\bibinfo {author} {\bibfnamefont {D.-S.}\ \bibnamefont
  {Han}}, \bibinfo {author} {\bibfnamefont {A.}~\bibnamefont {Vogel}}, \bibinfo
  {author} {\bibfnamefont {H.}~\bibnamefont {Jung}}, \bibinfo {author}
  {\bibfnamefont {K.-S.}\ \bibnamefont {Lee}}, \bibinfo {author} {\bibfnamefont
  {M.}~\bibnamefont {Weigand}}, \bibinfo {author} {\bibfnamefont
  {H.}~\bibnamefont {Stoll}}, \bibinfo {author} {\bibfnamefont
  {G.}~\bibnamefont {Sch\"{u}tz}}, \bibinfo {author} {\bibfnamefont
  {P.}~\bibnamefont {Fischer}}, \bibinfo {author} {\bibfnamefont
  {G.}~\bibnamefont {Meier}}, \ and\ \bibinfo {author} {\bibfnamefont {S.-K.}\
  \bibnamefont {Kim}},\ }\href@noop {} {\bibfield  {journal} {\bibinfo
  {journal} {Sci. Rep.}\ }\textbf {\bibinfo {volume} {3}},\ \bibinfo {pages}
  {2262} (\bibinfo {year} {2013})}\BibitemShut {NoStop}%
\bibitem [{\citenamefont {Belanovsky}\ \emph {et~al.}(2012)\citenamefont
  {Belanovsky}, \citenamefont {Locatelli}, \citenamefont {Skirdkov},
  \citenamefont {Abreu~Araujo}, \citenamefont {Grollier}, \citenamefont
  {Zvezdin}, \citenamefont {Cros},\ and\ \citenamefont
  {Zvezdin}}]{BelanovskyPRB2012}%
  \BibitemOpen
  \bibfield  {author} {\bibinfo {author} {\bibfnamefont {A.~D.}\ \bibnamefont
  {Belanovsky}}, \bibinfo {author} {\bibfnamefont {N.}~\bibnamefont
  {Locatelli}}, \bibinfo {author} {\bibfnamefont {P.~N.}\ \bibnamefont
  {Skirdkov}}, \bibinfo {author} {\bibfnamefont {F.}~\bibnamefont
  {Abreu~Araujo}}, \bibinfo {author} {\bibfnamefont {J.}~\bibnamefont
  {Grollier}}, \bibinfo {author} {\bibfnamefont {K.~A.}\ \bibnamefont
  {Zvezdin}}, \bibinfo {author} {\bibfnamefont {V.}~\bibnamefont {Cros}}, \
  and\ \bibinfo {author} {\bibfnamefont {A.~K.}\ \bibnamefont {Zvezdin}},\
  }\href@noop {} {\bibfield  {journal} {\bibinfo  {journal} {Phys. Rev. B}\
  }\textbf {\bibinfo {volume} {85}},\ \bibinfo {pages} {100409(R)} (\bibinfo
  {year} {2012})}\BibitemShut {NoStop}%
\bibitem [{\citenamefont {Belanovsky}\ \emph {et~al.}(2013)\citenamefont
  {Belanovsky}, \citenamefont {Locatelli}, \citenamefont {Skirdkov},
  \citenamefont {Abreu~Araujo}, \citenamefont {Zvezdin}, \citenamefont
  {Grollier}, \citenamefont {Cros},\ and\ \citenamefont
  {Zvezdin}}]{BelanovskyAPL2013}%
  \BibitemOpen
  \bibfield  {author} {\bibinfo {author} {\bibfnamefont {A.~D.}\ \bibnamefont
  {Belanovsky}}, \bibinfo {author} {\bibfnamefont {N.}~\bibnamefont
  {Locatelli}}, \bibinfo {author} {\bibfnamefont {P.~N.}\ \bibnamefont
  {Skirdkov}}, \bibinfo {author} {\bibfnamefont {F.}~\bibnamefont
  {Abreu~Araujo}}, \bibinfo {author} {\bibfnamefont {K.~A.}\ \bibnamefont
  {Zvezdin}}, \bibinfo {author} {\bibfnamefont {J.}~\bibnamefont {Grollier}},
  \bibinfo {author} {\bibfnamefont {V.}~\bibnamefont {Cros}}, \ and\ \bibinfo
  {author} {\bibfnamefont {A.~K.}\ \bibnamefont {Zvezdin}},\ }\href@noop {}
  {\bibfield  {journal} {\bibinfo  {journal} {Appl. Phys. Lett.}\ }\textbf
  {\bibinfo {volume} {103}},\ \bibinfo {pages} {122405} (\bibinfo {year}
  {2013})}\BibitemShut {NoStop}%
\bibitem [{\citenamefont {Abreu~Araujo}\ and\ \citenamefont
  {Grollier}(2016)}]{AbreuAraujoJAP2016}%
  \BibitemOpen
  \bibfield  {author} {\bibinfo {author} {\bibfnamefont {F.}~\bibnamefont
  {Abreu~Araujo}}\ and\ \bibinfo {author} {\bibfnamefont {J.}~\bibnamefont
  {Grollier}},\ }\href@noop {} {\bibfield  {journal} {\bibinfo  {journal} {J.
  Appl. Phys.}\ }\textbf {\bibinfo {volume} {120}},\ \bibinfo {pages} {103903}
  (\bibinfo {year} {2016})}\BibitemShut {NoStop}%
\bibitem [{\citenamefont {Abreu~Araujo}\ \emph {et~al.}(2015)\citenamefont
  {Abreu~Araujo}, \citenamefont {Belanovsky}, \citenamefont {Skirdkov},
  \citenamefont {Zvezdin}, \citenamefont {Zvezdin}, \citenamefont {Locatelli},
  \citenamefont {Lebrun}, \citenamefont {Grollier}, \citenamefont {Cros},
  \citenamefont {de~Loubens},\ and\ \citenamefont
  {Klein}}]{AbreuAraujoPRB2015}%
  \BibitemOpen
  \bibfield  {author} {\bibinfo {author} {\bibfnamefont {F.}~\bibnamefont
  {Abreu~Araujo}}, \bibinfo {author} {\bibfnamefont {A.~D.}\ \bibnamefont
  {Belanovsky}}, \bibinfo {author} {\bibfnamefont {P.~N.}\ \bibnamefont
  {Skirdkov}}, \bibinfo {author} {\bibfnamefont {K.~A.}\ \bibnamefont
  {Zvezdin}}, \bibinfo {author} {\bibfnamefont {A.~K.}\ \bibnamefont
  {Zvezdin}}, \bibinfo {author} {\bibfnamefont {N.}~\bibnamefont {Locatelli}},
  \bibinfo {author} {\bibfnamefont {R.}~\bibnamefont {Lebrun}}, \bibinfo
  {author} {\bibfnamefont {J.}~\bibnamefont {Grollier}}, \bibinfo {author}
  {\bibfnamefont {V.}~\bibnamefont {Cros}}, \bibinfo {author} {\bibfnamefont
  {G.}~\bibnamefont {de~Loubens}}, \ and\ \bibinfo {author} {\bibfnamefont
  {O.}~\bibnamefont {Klein}},\ }\href@noop {} {\bibfield  {journal} {\bibinfo
  {journal} {Phys. Rev. B}\ }\textbf {\bibinfo {volume} {92}},\ \bibinfo
  {pages} {045419} (\bibinfo {year} {2015})}\BibitemShut {NoStop}%
\bibitem [{\citenamefont {Dussaux}\ \emph {et~al.}(2012)\citenamefont
  {Dussaux}, \citenamefont {Khvalkovskiy}, \citenamefont {Bortolotti},
  \citenamefont {Grollier}, \citenamefont {Cros},\ and\ \citenamefont
  {Fert}}]{DussauxPRB2012}%
  \BibitemOpen
  \bibfield  {author} {\bibinfo {author} {\bibfnamefont {A.}~\bibnamefont
  {Dussaux}}, \bibinfo {author} {\bibfnamefont {A.~V.}\ \bibnamefont
  {Khvalkovskiy}}, \bibinfo {author} {\bibfnamefont {P.}~\bibnamefont
  {Bortolotti}}, \bibinfo {author} {\bibfnamefont {J.}~\bibnamefont
  {Grollier}}, \bibinfo {author} {\bibfnamefont {V.}~\bibnamefont {Cros}}, \
  and\ \bibinfo {author} {\bibfnamefont {A.}~\bibnamefont {Fert}},\ }\href@noop
  {} {\bibfield  {journal} {\bibinfo  {journal} {Phys. Rev. B}\ }\textbf
  {\bibinfo {volume} {86}},\ \bibinfo {pages} {014402} (\bibinfo {year}
  {2012})}\BibitemShut {NoStop}%
\bibitem [{\citenamefont {Flovik}\ \emph {et~al.}(2016)\citenamefont {Flovik},
  \citenamefont {Maci\`{a}},\ and\ \citenamefont
  {Wahlstr\"{o}m}}]{FlovikSREP2016}%
  \BibitemOpen
  \bibfield  {author} {\bibinfo {author} {\bibfnamefont {V.}~\bibnamefont
  {Flovik}}, \bibinfo {author} {\bibfnamefont {F.}~\bibnamefont {Maci\`{a}}}, \
  and\ \bibinfo {author} {\bibfnamefont {E.}~\bibnamefont {Wahlstr\"{o}m}},\
  }\href@noop {} {\bibfield  {journal} {\bibinfo  {journal} {Sci. Rep.}\
  }\textbf {\bibinfo {volume} {6}},\ \bibinfo {pages} {32528} (\bibinfo {year}
  {2016})}\BibitemShut {NoStop}%
\bibitem [{sup()}]{supplement}%
  \BibitemOpen
  \href@noop {} {\ }\bibinfo {note} {See the Supplemental Information for
  details.}\BibitemShut {Stop}%
\bibitem [{\citenamefont {Khvalkovskiy}\ \emph {et~al.}(2010)\citenamefont
  {Khvalkovskiy}, \citenamefont {Grollier}, \citenamefont {Locatelli},
  \citenamefont {Gorbunov}, \citenamefont {Zvezdin},\ and\ \citenamefont
  {Cros}}]{KhvalkovskiyAPL2010}%
  \BibitemOpen
  \bibfield  {author} {\bibinfo {author} {\bibfnamefont {A.~V.}\ \bibnamefont
  {Khvalkovskiy}}, \bibinfo {author} {\bibfnamefont {J.}~\bibnamefont
  {Grollier}}, \bibinfo {author} {\bibfnamefont {N.}~\bibnamefont {Locatelli}},
  \bibinfo {author} {\bibfnamefont {Y.~V.}\ \bibnamefont {Gorbunov}}, \bibinfo
  {author} {\bibfnamefont {K.~A.}\ \bibnamefont {Zvezdin}}, \ and\ \bibinfo
  {author} {\bibfnamefont {V.}~\bibnamefont {Cros}},\ }\href@noop {} {\bibfield
   {journal} {\bibinfo  {journal} {Appl. Phys. Lett.}\ }\textbf {\bibinfo
  {volume} {96}},\ \bibinfo {pages} {212507} (\bibinfo {year}
  {2010})}\BibitemShut {NoStop}%
\bibitem [{\citenamefont {Guslienko}\ \emph {et~al.}(2002)\citenamefont
  {Guslienko}, \citenamefont {Ivanov}, \citenamefont {Novosad}, \citenamefont
  {Otani}, \citenamefont {Shima},\ and\ \citenamefont
  {Fukamichi}}]{GuslienkoJAP2002}%
  \BibitemOpen
  \bibfield  {author} {\bibinfo {author} {\bibfnamefont {K.~Y.}\ \bibnamefont
  {Guslienko}}, \bibinfo {author} {\bibfnamefont {B.~A.}\ \bibnamefont
  {Ivanov}}, \bibinfo {author} {\bibfnamefont {V.}~\bibnamefont {Novosad}},
  \bibinfo {author} {\bibfnamefont {Y.}~\bibnamefont {Otani}}, \bibinfo
  {author} {\bibfnamefont {H.}~\bibnamefont {Shima}}, \ and\ \bibinfo {author}
  {\bibfnamefont {K.}~\bibnamefont {Fukamichi}},\ }\href@noop {} {\bibfield
  {journal} {\bibinfo  {journal} {J. Appl. Phys.}\ }\textbf {\bibinfo {volume}
  {91}},\ \bibinfo {pages} {8037} (\bibinfo {year} {2002})}\BibitemShut
  {NoStop}%
\bibitem [{\citenamefont {Naletov}\ \emph {et~al.}(2011)\citenamefont
  {Naletov}, \citenamefont {de~Loubens}, \citenamefont {Albuquerque},
  \citenamefont {Borlenghi}, \citenamefont {Cros}, \citenamefont {Faini},
  \citenamefont {Grollier}, \citenamefont {Hurdequint}, \citenamefont
  {Locatelli}, \citenamefont {Pigeau}, \citenamefont {Slavin}, \citenamefont
  {Tiberkevich}, \citenamefont {Ulysse}, \citenamefont {Valet},\ and\
  \citenamefont {Klein}}]{NaletovPRB2011}%
  \BibitemOpen
  \bibfield  {author} {\bibinfo {author} {\bibfnamefont {V.~V.}\ \bibnamefont
  {Naletov}}, \bibinfo {author} {\bibfnamefont {G.}~\bibnamefont {de~Loubens}},
  \bibinfo {author} {\bibfnamefont {G.}~\bibnamefont {Albuquerque}}, \bibinfo
  {author} {\bibfnamefont {S.}~\bibnamefont {Borlenghi}}, \bibinfo {author}
  {\bibfnamefont {V.}~\bibnamefont {Cros}}, \bibinfo {author} {\bibfnamefont
  {G.}~\bibnamefont {Faini}}, \bibinfo {author} {\bibfnamefont
  {J.}~\bibnamefont {Grollier}}, \bibinfo {author} {\bibfnamefont
  {H.}~\bibnamefont {Hurdequint}}, \bibinfo {author} {\bibfnamefont
  {N.}~\bibnamefont {Locatelli}}, \bibinfo {author} {\bibfnamefont
  {B.}~\bibnamefont {Pigeau}}, \bibinfo {author} {\bibfnamefont {A.~N.}\
  \bibnamefont {Slavin}}, \bibinfo {author} {\bibfnamefont {V.~S.}\
  \bibnamefont {Tiberkevich}}, \bibinfo {author} {\bibfnamefont
  {C.}~\bibnamefont {Ulysse}}, \bibinfo {author} {\bibfnamefont
  {T.}~\bibnamefont {Valet}}, \ and\ \bibinfo {author} {\bibfnamefont
  {O.}~\bibnamefont {Klein}},\ }\href {\doibase 10.1103/PhysRevB.84.224423}
  {\bibfield  {journal} {\bibinfo  {journal} {Phys. Rev. B}\ }\textbf {\bibinfo
  {volume} {84}},\ \bibinfo {pages} {224423} (\bibinfo {year}
  {2011})}\BibitemShut {NoStop}%
\bibitem [{\citenamefont {Guslienko}(2006)}]{GuslienkoAPL2006}%
  \BibitemOpen
  \bibfield  {author} {\bibinfo {author} {\bibfnamefont {K.~Y.}\ \bibnamefont
  {Guslienko}},\ }\href@noop {} {\bibfield  {journal} {\bibinfo  {journal}
  {Appl. Phys. Lett.}\ }\textbf {\bibinfo {volume} {89}},\ \bibinfo {pages}
  {022510} (\bibinfo {year} {2006})}\BibitemShut {NoStop}%
\bibitem [{\citenamefont {Ivanov}\ and\ \citenamefont
  {Zaspel}(2007)}]{IvanovPRL2007}%
  \BibitemOpen
  \bibfield  {author} {\bibinfo {author} {\bibfnamefont {B.~A.}\ \bibnamefont
  {Ivanov}}\ and\ \bibinfo {author} {\bibfnamefont {C.~E.}\ \bibnamefont
  {Zaspel}},\ }\href@noop {} {\bibfield  {journal} {\bibinfo  {journal} {Phys.
  Rev. Lett.}\ }\textbf {\bibinfo {volume} {99}},\ \bibinfo {pages} {247208}
  (\bibinfo {year} {2007})}\BibitemShut {NoStop}%
\bibitem [{\citenamefont {Khvalkovskiy}\ \emph {et~al.}(2009)\citenamefont
  {Khvalkovskiy}, \citenamefont {Grollier}, \citenamefont {Dussaux},
  \citenamefont {Zvezdin},\ and\ \citenamefont {Cros}}]{KhvalkovskiyPRB2009}%
  \BibitemOpen
  \bibfield  {author} {\bibinfo {author} {\bibfnamefont {A.~V.}\ \bibnamefont
  {Khvalkovskiy}}, \bibinfo {author} {\bibfnamefont {J.}~\bibnamefont
  {Grollier}}, \bibinfo {author} {\bibfnamefont {A.}~\bibnamefont {Dussaux}},
  \bibinfo {author} {\bibfnamefont {K.~A.}\ \bibnamefont {Zvezdin}}, \ and\
  \bibinfo {author} {\bibfnamefont {V.}~\bibnamefont {Cros}},\ }\href@noop {}
  {\bibfield  {journal} {\bibinfo  {journal} {Phys. Rev. B}\ }\textbf {\bibinfo
  {volume} {80}},\ \bibinfo {pages} {140401(R)} (\bibinfo {year}
  {2009})}\BibitemShut {NoStop}%
\bibitem [{\citenamefont {Adler}(1973)}]{AdlerIEEE1973}%
  \BibitemOpen
  \bibfield  {author} {\bibinfo {author} {\bibfnamefont {R.}~\bibnamefont
  {Adler}},\ }\href@noop {} {\bibfield  {journal} {\bibinfo  {journal} {Proc.
  IRE}\ }\textbf {\bibinfo {volume} {34}},\ \bibinfo {pages} {351} (\bibinfo
  {year} {1973})}\BibitemShut {NoStop}%
\bibitem [{\citenamefont {Slavin}\ and\ \citenamefont
  {Tiberkevich}(2006)}]{SlavinPRB2006}%
  \BibitemOpen
  \bibfield  {author} {\bibinfo {author} {\bibfnamefont {A.~N.}\ \bibnamefont
  {Slavin}}\ and\ \bibinfo {author} {\bibfnamefont {V.~S.}\ \bibnamefont
  {Tiberkevich}},\ }\href@noop {} {\bibfield  {journal} {\bibinfo  {journal}
  {Phys. Rev. B}\ }\textbf {\bibinfo {volume} {74}},\ \bibinfo {pages} {104401}
  (\bibinfo {year} {2006})}\BibitemShut {NoStop}%
\bibitem [{\citenamefont {Hong}\ and\ \citenamefont
  {Strogatz}(2011)}]{HongPRL2011}%
  \BibitemOpen
  \bibfield  {author} {\bibinfo {author} {\bibfnamefont {H.}~\bibnamefont
  {Hong}}\ and\ \bibinfo {author} {\bibfnamefont {S.~H.}\ \bibnamefont
  {Strogatz}},\ }\href@noop {} {\bibfield  {journal} {\bibinfo  {journal}
  {Phys. Rev. Lett.}\ }\textbf {\bibinfo {volume} {106}},\ \bibinfo {pages}
  {054102} (\bibinfo {year} {2011})}\BibitemShut {NoStop}%
\bibitem [{\citenamefont {Zhou}\ \emph {et~al.}(2007)\citenamefont {Zhou},
  \citenamefont {Persson},\ and\ \citenamefont {\r{A}kerman}}]{ZhouJAP2007}%
  \BibitemOpen
  \bibfield  {author} {\bibinfo {author} {\bibfnamefont {Y.}~\bibnamefont
  {Zhou}}, \bibinfo {author} {\bibfnamefont {J.}~\bibnamefont {Persson}}, \
  and\ \bibinfo {author} {\bibfnamefont {J.}~\bibnamefont {\r{A}kerman}},\
  }\href@noop {} {\bibfield  {journal} {\bibinfo  {journal} {J. Appl. Phys.}\
  }\textbf {\bibinfo {volume} {101}},\ \bibinfo {pages} {09A510} (\bibinfo
  {year} {2007})}\BibitemShut {NoStop}%
\end{thebibliography}
\end{document}


\title{Supplemental Information to "Probing phase coupling between two spin-torque nano-oscillators with an external source"}


\author{Yi Li}
\email{yi.li@cea.fr}
\affiliation{Service de Physique de l'\'{E}tat Condens\'{e}, CEA, CNRS, Universit\'{e} Paris-Saclay, Gif-sur-Yvette, France}

\author{Xavier de Milly}
\affiliation{Service de Physique de l'\'{E}tat Condens\'{e}, CEA, CNRS, Universit\'{e} Paris-Saclay, Gif-sur-Yvette, France}

\author{Flavio Abreu Araujo}
\affiliation{Unit\'{e} Mixte de Physique CNRS, Thales, Univ. Paris-Sud, Universit\'{e} Paris-Saclay, Palaiseau, France}

\author{Olivier Klein}
\affiliation{SPINTEC, Univ. Grenoble Alpes / CEA / CNRS, 38000 Grenoble, France}

\author{Vincent Cros}
\affiliation{Unit\'{e} Mixte de Physique CNRS, Thales, Univ. Paris-Sud, Universit\'{e} Paris-Saclay, Palaiseau, France}

\author{Julie Grollier}
\affiliation{Unit\'{e} Mixte de Physique CNRS, Thales, Univ. Paris-Sud, Universit\'{e} Paris-Saclay, Palaiseau, France}

\author{Gr\'{e}goire de Loubens}
\email{gregoire.deloubens@cea.fr}
\affiliation{Service de Physique de l'\'{E}tat Condens\'{e}, CEA, CNRS, Universit\'{e} Paris-Saclay, Gif-sur-Yvette, France}

\date{\today}

\maketitle

\subsection{Chirality states measured by magnetoresistance}

\begin{figure}[htb]
 \centering
 \includegraphics[width=4.0 in]{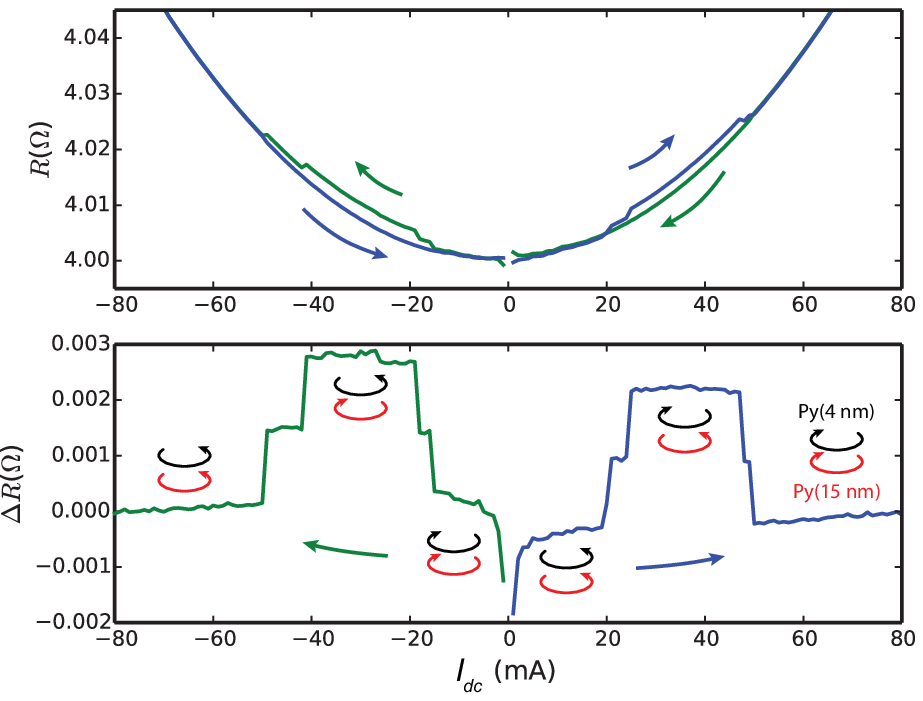}
 \caption{(a) Hysteresis loop of $R$ as a function of current. The arrows show the direction of $I_{dc}$ evolutions. (b) $\Delta R$ of the hysteresis as a function of current. The red and black arrows denote the chirality states of Py(15 nm) and Py(4 nm) vortices, respectively. The chirality states are the same for the two STVOs at the highest or lowest stage of $\Delta R$.}
 \label{fig1}
\end{figure}

The chiralities of the four vortices in the pair of STVO (two for each STVO, one in the thick Py(15 nm) disc and one from the thin Py(4 nm) disc), which define the curling direction of the magnetization around vortex core, can be accessed by the magnetoresistance effect. Figure \ref{fig1}(a) shows the hysteresis loop of electrical resistance ($R$) of the sample as a function of current ($I_{dc}$) at zero biasing field. The resistance difference $\Delta R$ between the upper and lower branches is extracted in Figure \ref{fig1}(b). When the absolute value of $I_{dc}$, or $|I_{dc}|$, is above 60 mA, the chiralities of the two Py layers in each STVO are the same, both clockwise for $I_{dc}>0$ and counter-clockwise for $I_{dc}<0$. The magnetoresistance is minimized due to parallel magnetizations in each STVO. By reducing $|I_{dc}|$ and reversing its sign, two abrupt resistance increases are firstly observed when $|I_{dc}|$ is around 20 mA corresponding to the chirality switchings of the two Py(4 nm) layers. Keeping increasing $|I_{dc}|$, two additional jumps can also be observed when $|I_{dc}|$ is between 40 and 50 mA corresponding to the chirality switchings of the two Py(15 nm) layers.

Owing to the similar chirality switching currents and amplitudes of resistance jump, the data show that the two nanopillar spin valves are close in geometry as designed. It also shows that at the working current of $I_{dc}=95$ mA, the four chirality states are ensured to be the same.

\subsection{From the Thiele equation to the oscillator phase equation}

\begin{figure}[htb]
 \centering
 \includegraphics[width=3.0 in]{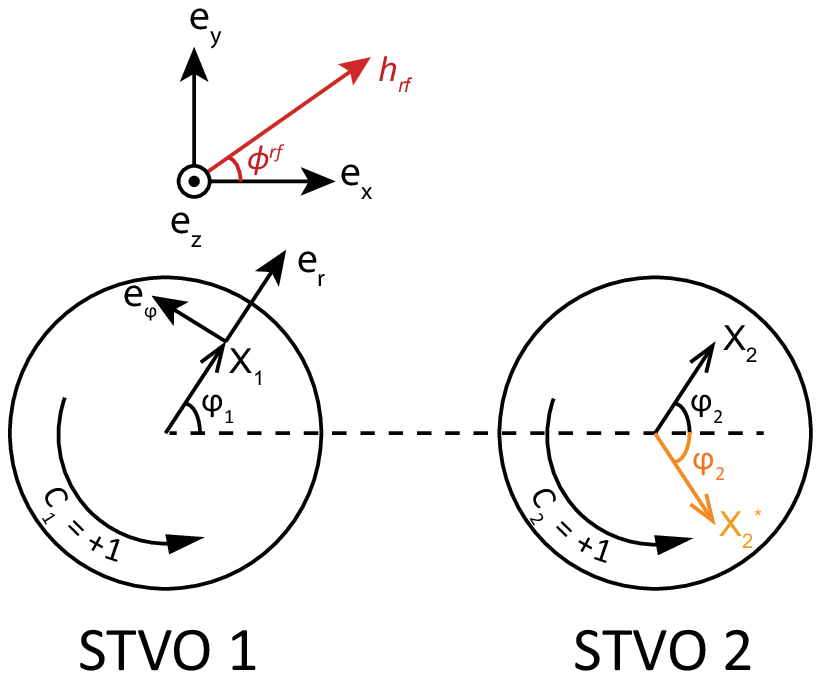}
 \caption{Coordinate system for two coupled STVOs. $\mathbf{e}_r$ and $\mathbf{e}_\varphi$ are referred to the STVO in question.}
 \label{fig2}
\end{figure}

Here we briefly show how to obtain the oscillator phase equations (Eq. 2 of the main text) from the Thiele equation of a magnetic vortex\cite{IvanovPRL2007,KhvalkovskiyPRB2009}. We write out the Thiele equation for STVO 1:
\begin{equation}
 (-G_1\mathbf{e}_z) \times \mathbf{\dot{X}}_1 - \kappa_1\mathbf{X}_1 - D_1 \mathbf{\dot{X}}_1 + \mathbf{F}_\text{1,STT} + \mathbf{F}_{1e} + \mathbf{F}_{1d} = 0
\label{eq01}
\end{equation}
where $\mathbf{X_1}$ is the position of the vortex core, $G_1=2\pi p_1\mu_0M_{1s} h/\gamma$ is the amplitude of gyrovector, $\kappa_1$ denotes the stiffness of the vortex confining force, $D_1$ is the damping term proportional to the Gilbert damping, $\mathbf{F}_\text{1,STT}=\sigma_1 J(\mathbf{e}_z \times \mathbf{X}_1)$ is the anti-damping term from spin-transfer torque, $\mathbf{F}_{1e}$ is the force from microwave field coupling and $\mathbf{F}_{1d}$ is the force from dipolar coupling. In the expression of $G_1$ the vortex polarity $p_1$ determines the sign of angular frequency, \textit{i. e.} the direction of vortex core gyration. Here we set both chiralities of the two STVOs to be counter-clockwise from top view ($C_{1,2}=+1$). Because the two STVOs are patterned from the same film stack, we take $|G_1|=|G_2|=G$ for the two STVOs so that $G_{1,2} = p_{1,2}G$. The coordinate system is defined in Fig. \ref{fig2}.

In Eq. (\ref{eq01}) $\mathbf{F}_{1e}$ and $\mathbf{F}_{1d}$ can be calculated from the energy expression as $\mathbf{F}_{1e,d} = -\partial W_{1e,d} / \partial \mathbf{X_1}$. In vortex state, the coupling energy to the microwave field can be expressed as $W_{1e} = -\mathbf{h}_{rf} \cdot (\mu_0 m_1\mathbf{X_1} \times C_1\mathbf{e}_z) $, where $\mathbf{h}_{rf}$ is the rf field, $C_1$ is the chirality ($C_1=+1$ denotes counter-clockwise) and $m_1\sim (2/3)M_{1s}V_1/R_1$ is the effective dynamic magnetization divided by the disc radius\cite{GuslienkoJAP2002}. This yields:
\begin{equation}
\mathbf{F}_{1e} = C_1\mu_0m_1(\mathbf{e}_z \times\mathbf{h}_{rf})
\label{eq02}
\end{equation}

The dipolar coupling energy can be expressed as\cite{AbreuAraujoPRB2015} $W_{1d}=-\mu^\text{P} \mathbf{X_1} \cdot \mathbf{X_2} + \mu^\text{AP} \mathbf{X_1} \cdot \mathbf{X_2^*}$, with $\mu^\text{AP}=3\mu^\text{P}$. Here $\mathbf{X_2^*}$ is the conjugate of $\mathbf{X_2}$ (Fig. \ref{fig2}). Thus the dipolar coupling force can be expressed as:
\begin{equation}
\mathbf{F}_{1d} = \mu^\text{P} \mathbf{X_2} - \mu^\text{AP} \mathbf{X_2^*}
\label{eq03}
\end{equation}

In polar coordinates we assume that the amplitudes and phases of $\mathbf{X}_{1,2}$ to be $X_{1,2}$ and $\varphi_{1,2}$. If we consider only the steady-orbit solution, by setting $dX_{1,2}/dt = 0$ and taking the dot product of Eq. (\ref{eq01}) and $\mathbf{e}_r$, we obtain:
\begin{equation}
(p_1{d\varphi_1 \over dt} - \omega_{1g})GX_1 + \mathbf{F}_{1e} \cdot \mathbf{e}_r + \mathbf{F}_{1d} \cdot \mathbf{e}_r = 0
\label{eq04}
\end{equation}
with $\omega_{1g} = \kappa_1/G$. We will show below that Eq. (\ref{eq04}) can be written in a generalized form, presented as Eq. (2) of the main text:
\begin{equation}
-{d\theta_i \over dt} +\delta_{i} +\Omega_{j i}\cos(\theta_i+\gamma_i^{NL} - \theta_j) - \Delta_{i e}\sin(\theta_i+\gamma_i^{NL}+\gamma_i^{rf}) = 0
\label{eq05}
\end{equation}
where $\theta_i = \omega_e t  - p_i \varphi_i$ is the phase difference between the microwave and the oscillator, $\delta_{i}=\omega_e - \omega_{i g}$ is the frequency mismatch, $\Omega_{j i} = \Omega(X_j / X_i)$ is the effective dipolar coupling strength, $\Delta_{i e}$ is the microwave coupling strength, $\gamma_i^{NL}$ is the intrinsic phase shift due to nonlinearity, and $\gamma_i^{rf}$ is the microwave coupling phase determined by the geometric alignment of the microwave field to the oscillator. The index is defined as $(i, j)=(1, 2)$ or $(2, 1)$.

\subsubsection{Pure phase oscillators, $\gamma_i^{NL}=0$}

If the angular frequency $\omega_{1g}$ does not depend on $X_1$, the nonlinearity of the STVO is zero and the nonlinear phase shift $\gamma_i^{NL}=0$ in Eq. (\ref{eq05}).

\textbf{Parallel polarity alignment:} we take the polarities as $p_1=p_2=+1$. For $\mathbf{F}_{1e}$ we assume an in-plane linearly polarized microwave field with an arbitrary direction $\mathbf{h}_{rf}=h_{rf}(\cos\phi^{rf}\mathbf{e}_x+\sin\phi^{rf}\mathbf{e}_y) \cos(\omega_e t)$, as shown in Fig. \ref{fig2}. Using Eq. (\ref{eq02}), the microwave term $\mathbf{F}_{1e} \cdot \mathbf{e}_r$ in Eq. (\ref{eq04}) can be expressed as:
\begin{align}
\mathbf{F}_{1e} \cdot \mathbf{e}_r &= f_{1e} (\cos\phi^{rf}\mathbf{e}_y \cdot \mathbf{e}_r-\sin\phi^{rf}\mathbf{e}_x \cdot \mathbf{e}_r) \cos(\omega_e t) \nonumber \\
&= f_{1e} \sin(\varphi_1-\phi^{rf}) \cos(\omega_e t) \nonumber \\
&= (f_{1e}/2) [\sin(\omega_e t+\varphi_1 -\phi^{rf}) - \sin(\omega_e t-\varphi_1 +\phi^{rf})]
\label{eq06}
\end{align}
with $f_{1e} = C_1\mu_0 m_1 h_{rf}$. The two terms in Eq. (\ref{eq06}) are the clockwise and counter-clockwise polarized components. In positive polarity state, the first term serves as a high frequency component with zero time average so we will neglect it.

For the dipolar coupling term, we have:
\begin{align}
\mathbf{F}_{1d} \cdot \mathbf{e}_r &=  \mu^\text{P} (\mathbf{X_2} \cdot \mathbf{e}_r)  - \mu^\text{AP} (\mathbf{X_2^*} \cdot \mathbf{e}_r ) \nonumber \\
&= \mu^\text{P} X_2 \cos(\varphi_2 - \varphi_1) - \mu^\text{AP} X_2 \cos(\varphi_2 + \varphi_1)
\label{eq07}
\end{align}
In the parallel polarity state the first term mainly contributes to Eq. (\ref{eq04}) and the second term is a high-frequency term with an average of zero. As a result, Eq. (\ref{eq04}) becomes:
\begin{equation}
{d\varphi_1 \over dt} - \omega_{1g} +{ \mu^\text{P}\over G} {X_2 \over X_1}\cos(\varphi_2 - \varphi_1) - {f_{1e} \over 2GX_1}\sin(\omega_e t-\varphi_1+\phi^{rf}) = 0
\label{eq08}
\end{equation}
Comparing with Eq. (\ref{eq05}), we find $\Omega={ \mu^\text{P}\over G}$, $\Delta_{1e} = {f_{1e} \over 2GX_1}$ and $\gamma_1^{rf}=\phi^{rf}$. The equation for STVO 2 is the same except that the indices 1 and 2 are exchanged.

\textbf{Antiparallel polarity alignment:} we take the polarities as $p_1=+1$ and $p_2=-1$. This time $\varphi_1$ will increase but $\varphi_2$ will decrease with time. In Eq. (\ref{eq06}) we still take the second term for STVO 1. In Eq. (\ref{eq07}), however, the $\cos(\varphi_2 + \varphi_1)$ term mainly contributes to the phase interaction and the first term is a high frequency term. Eq. (\ref{eq04}) becomes:
\begin{equation}
{d\varphi_1 \over dt} - \omega_{1g} - { \mu^\text{AP}\over G} {X_2 \over X_1}\cos(\varphi_2 + \varphi_1) - {f_{1e} \over 2GX_1}\sin(\omega_e t-\varphi_1+\phi^{rf}) = 0
\label{eq09}
\end{equation}
Comparing with Eq. (\ref{eq05}), we find $\Omega=-{ \mu^\text{AP}\over G}$, $\Delta_{1e} = {f_{1e} \over 2GX_1}$ and $\gamma_1^{rf}=\phi^{rf}$.

For STVO 2, the active term in Eq. (\ref{eq06}) is changed to the first term. Along with the negative polarity, the oscillator equation is:
\begin{equation}
-{d\varphi_2 \over dt} - \omega_{2g} - { \mu^\text{AP}\over G} {X_1 \over X_2}\cos(\varphi_2 + \varphi_1) + {f_{2e} \over 2GX_2}\sin(\omega_e t + \varphi_2 -\phi^{rf}) = 0
\label{eq10}
\end{equation}
Comparing with Eq. (\ref{eq05}), we find $\Omega=-{ \mu^\text{AP}\over G}$, $\Delta_{2e} = {f_{2e} \over 2GX_2}$ and $\gamma_2^{rf}=\pi-\phi^{rf}$.

For $\gamma_i^{rf}$ the different sign in $\phi^{rf}$ is due to the polarity change, with which the vortex gyrates along different circular directions. The vortex dynamics thus couples to different polarizations of the linearly polarized microwave field, as shown in Eq. (\ref{eq06})

The additional phase of $\pi$ in $\gamma_2^{rf}$ is due to the cross product in Eq. (\ref{eq02}), adding a relative phase of $\pi/2$ to $\gamma_1^{rf}$ and $-\pi/2$ to $\gamma_2^{rf}$.


\subsubsection{Coupled phase-amplitude oscillators, $\gamma_i^{NL}\neq 0$}

Here we include finite nonlinearity, which couples the phase equation to the amplitude equation derived from Eq. (\ref{eq01}). Two effects are considered\cite{SlavinIEEE2009}. First, the auto-oscillation frequency depends on the precession amplitude, as $\omega=\omega_0 + N p$ where $p=|\mathbf{X}|^2$ and $N$ is the nonlinear frequency shift. Second, the relaxation rates also depend on the precession amplitude, with the first-order deviations defined as $G_\pm = \partial \Gamma_\pm/\partial p$ where $\Gamma_+=D\omega_g/G$ and $\Gamma_-=\sigma J/G$ from Eq. (\ref{eq01}). This allows to define a dimensionless nonlinearity coefficient $\nu = N / (G_+-G_-)$. As a result, a factor of $\sqrt{1+\nu^2}$ will be multiplied to all the coupling coefficients, and a phase shift of $\gamma_i^{NL}=\tan^{-1}\nu$ will be added as shown in Eq. (\ref{eq05}). We note that finite damping will also add a small correction, as derived in the Supplemental Information of Ref. \cite{locatelliSREP2015}.

We emphasize that in Eq. (\ref{eq05}), the finite phase shift $\gamma_i^{NL}$ of the dipolar coupling term has the same value as the phase shift $\gamma_i^{NL}$ of the microwave coupling term, because they both come from the nonlinearity of the $i$th STVO.

\subsection{The limit of weak microwave coupling}

In the theoretical derivation of the main text, the limit of weak microwave coupling (i.e. microwave field) is required in order to obtain a simple expression for $\theta_1$ when oscillator 1 is phase-locked to the microwave field. In this situation, because the frequency mismatch between the microwave field and oscillator 2 is much greater than the coupling strength $\Delta_{2e}$, the phase of oscillator 2 will act as if there is no frequency pulling from the microwave field. The time evolution of the term $\omega_{1g} - \Omega_{21}\cos(\theta_1 + \gamma_1^{NL}-\theta_2)$ in Eq. (2) of the main text can be replaced by its time average, $\omega_{1g}^c$, the frequency of oscillator 1 in the presence of dipolar coupling which corresponds to the experimentally measured auto-oscillation peak without microwave field. The phase-locked state of $\theta_1$ can be analytically described by:
\begin{equation}
\sin(\theta_1+\gamma_1^{NL}+\gamma_1^{rf}) = {\omega_e-\omega^c_{1g} \over \Delta_{1e} }
\label{eq11}
\end{equation}
In Figs. 2 and 3 of the main text, this approximation also means that the remote frequency pullings (the relative frequency shift of oscillator 2 when oscillator 1 is phase-locked to microwave field, or \textit{vise versa}) are much smaller than the frequency mismatch of the two oscillators.

However, if the microwave coupling is not weak, the term $\omega^c_{1g}$ will be a function of $\theta_2$, which again depends on $\theta_1$. In this situation numerical simulations are required to solve the frequency dependence of $\theta_1$. Furthermore, with strong microwave disturbance, the term $\Omega_{21}\cos(\theta_1 + \gamma_1^{NL}-\theta_2)$  in Eq. (2) of the main text may not be treated as a ``time-averaged'' constant, because the term $\omega^c_{1g}$ in Eq. (\ref{eq11}) strongly varies with time. In this situation we cannot even reach a ``phase-locked'' state, but instead a complicated beating state between the two oscillators and the microwave field.

\subsection{Data fitting}

In this section we would like to specify how we take the parameters as inputs in order to extract the values in the Table I of the manuscript. As discussed in the main text, we fit the peak values $\omega_{1,2}$ as a function of microwave frequency $\omega_e$, described in the Eq. (4) of the main text. The fit parameters are $\Omega$ and $\gamma_i^{NL}$. To do so we need to know a few parameters in the equation.

\textbf{Values of $X_j/X_i$:} The amplitude ratio modifies the effective dipolar coupling strength $\Omega_{ji}=(X_j/X_i)\Omega$. It is reflected by the asymmetry of the frequency pullings between STVOs 1 and 2. Because the two STVOs have identical nominal geometry, their auto-oscillation amplitude should be similar. In the data analysis we set $X_1/X_2$ to be 1.1 for the antiparallel polarity alignment (Figs. 3 a-c of the main text) and 1.0 for the parallel polarity alignment (Figs. 3d of the main text) to account for slightly different frequency pulling in the two reciprocal cases of each figure.

\textbf{Values of $\omega_{ig}$ and $\omega_{ig}^c$:} The term $\omega_{ig}^c$ has been introduced in Section C of the Supplemental Information. In the data analysis we take it to be the same as the free-running auto-oscillation frequency $\omega_{ig}$ when the microwave field frequency is far away. The parameter $\omega_{ig}$ is defined as the free-running auto-oscillation frequency, which is in principle different from $\omega_{ig}^c$. However since the dipolar-coupling frequency pulling before mutual synchronization is weak in Fig. 1(g) of the main text, the value of $\omega_{ig}$ should be very close to $\omega_{ig}^c$. In this case we take $\omega_{ig}$ to be equal to $\omega_{ig}^c$.

\textbf{Values of $\gamma_i^{rf}$:} In the data fitting, $\gamma_2^{rf}-\gamma_1^{rf}$ is used as an input parameter. In our data analysis, its value is determined by the geometric
alignment of the microwave field to the two oscillators. In Fig. 1(a) of the main text, the microwave field $\mathbf{h_{rf}}$ is parallel to the alignment of the two STVOs, corresponding to $\phi^{rf}=0$ in Fig. \ref{fig2}. As a result, in the case of parallel polarity alignment, $\gamma_1^{rf}=\gamma_2^{rf}=\phi^{rf}$, thus $\gamma_2^{rf}-\gamma_1^{rf}=0$; in the case of antiparallel polarity alignment, $\gamma_2^{rf}-\gamma_1^{rf}=\pi - 2\phi^{rf}=\pi$. Nevertheless the large phase shift $\gamma^{NL}_i$ measured from the experiments might suggest that a finite $\phi^{rf}$ exists due to parasitic rf couplings between the antenna and sample circuits.